\documentclass[%
preprint,
 amsmath,amssymb,
 aps,
]{revtex4-2}

\usepackage{microtype}
\usepackage{comment}
\usepackage{natbib}
\usepackage{graphicx}
\usepackage{caption}
\usepackage{ragged2e}
\DeclareCaptionJustification{myjustified}{\justifying}
\usepackage{dcolumn}
\usepackage{bm}
\usepackage{empheq}
\usepackage{makecell}
\usepackage{adjustbox}
\usepackage{booktabs}
\usepackage{tabularx}
\usepackage{xcolor}
\usepackage{array}
\usepackage{rotating}
\usepackage{placeins}
\usepackage{multirow}
\usepackage[hidelinks]{hyperref}

\makeatletter
\renewcommand{\bibsection}{%
  \section*{References}%
}
\makeatother

\begin{document}


\title{On the capacitance gradient description \\
in Heterodyne Kelvin Probe Force Microscopy}

\author{
Hugo Valloire$^{(1)}$, Sylvain Clair$^{(1)}$, Christian Loppacher$^{(1)}$ and Laurent Nony$^{(1),\dagger}$ \\
Benjamin Gr\'evin$^{(2),\ddagger}$\\
{\scriptsize $^\dagger$\texttt{laurent.nony@im2np.fr}}\\
{\scriptsize $^\ddagger$\texttt{benjamin.grevin@neel.cnrs.fr}}
}

\affiliation{%
\vspace{0.5em}%
\begin{tabular}{c}
$^{(1)}$ Aix Marseille Université, CNRS, Univ Toulon, IM2NP, Marseille, France \\[0.25em]
$^{(2)}$ Institut Néel, UPR 2940 CNRS, Grenoble, France
\end{tabular}%
\vspace{0.5em}%
}

\date{\today}

\begin{abstract}
Kelvin probe force microscopy (KPFM) is a highly sensitive technique for probing local surface-potential variations. In KPFM, the electrostatic force between a conductive tip and the surface is proportional to the square of their potential difference and to the tip-surface capacitance gradient. In heterodyne KPFM, the tip oscillates above the surface. The electrostatic interaction is therefore commonly described by combining the tip-surface bias-modulated electric field with a first-order truncated Taylor-series expansion of the capacitance gradient. This treatment is analytically convenient, but it restricts the formalism to a poorly defined small-oscillation-amplitude regime and leaves unresolved the question of convergence of the underlying series expansion. Here, a rigorous spectral description of the capacitance-gradient dynamics and of the resulting electrostatic force is established, valid beyond this approximation. A general non-truncated Taylor-series description of the capacitance gradient is first formulated, and its convergence is established for a realistic capacitance model, independently of whether the cantilever motion is monomodal or bimodal. Then, it is shown in the monomodal case that the capacitance gradient can be described equivalently by Fourier-series and Taylor-series expansions. Explicit expressions for the dominant Fourier coefficients are obtained, and term-significance-based order-truncation-regime criteria are provided to replace the usual qualitative notion of ``small-oscillation-amplitude regime''. The same formalism is applied to bimodal motion, and the effective capacitance-gradient coefficients governing the static, first-eigenmode, and second-eigenmode components of the electrostatic interaction are derived. The convergence of the Taylor-based coefficients toward the Fourier coefficients is confirmed by numerical simulations, and the resulting term-significance-based order-truncation-regime hierarchy is supported in both monomodal and bimodal configurations. This work establishes the formal foundation required to describe electrostatic force components and AFM observables in open-loop heterodyne KPFM experiments.
\end{abstract}

\maketitle

\clearpage

\section{Introduction}\label{sec:introduction}

Kelvin probe force microscopy (KPFM) is a well-established electrostatic variant of atomic force microscopy (AFM) that probes the electrostatic landscape of a sample surface with nanometer resolution \cite{Melitz2011a,Sadewasser2012a,Sadewasser2018a}. It measures the tip-surface contact potential difference (CPD, denoted $V_{\mathrm{cpd}}$). This quantity stems from the tip-surface work-function difference and from electric charges or static dipoles in the system under consideration \cite{Kelvin1898a,Zisman1932a}. The operating principle of KPFM is to minimize the tip-surface electrostatic force by applying a DC compensation bias that matches the CPD. The tip and sample surface form a capacitance that, upon biasing, gives rise to an electrostatic force that depends on both time and tip-surface distance and influences the cantilever deflection. CPD compensation is achieved by combining a lock-in amplifier (LIA) with a DC compensation-bias controller, the so-called KPFM controller. The LIA output provides an AC bias-modulation voltage that is added to the DC voltage applied to the tip or sample. The resulting electrostatic force, $F_{\mathrm{el}}(z(t),t)$, is proportional to the tip-surface capacitance gradient (CG) and to the square of the instantaneous tip-surface potential difference:

\begin{subequations}\label{eq:Fel}
\begin{empheq}[left=\empheqlbrace]{align}
F_{\mathrm{el}}(z(t),t) &=
\frac{1}{2}\frac{dC(z)}{dz}\left[ V_{\mathrm{DC}} - V_{\mathrm{cpd}} + V_{\mathrm{mod}}(t) \right]^2
\label{eq:Fel_a} \\
V_{\mathrm{mod}}(t) &=
U_{\mathrm{mod}}\cos\left( \omega_{\mathrm{mod}} t + \Phi_{\mathrm{mod}} \right),
\label{eq:Fel_b}
\end{empheq}
\end{subequations}

\noindent where $V_{\mathrm{DC}}$, $V_{\mathrm{mod}}(t)$, and $V_{\mathrm{cpd}}$ denote the DC bias, the AC bias-modulation voltage, and the tip-surface CPD, respectively \cite{CG_Note_01}. In closed-loop KPFM, $V_{\mathrm{DC}}$ acts as the compensation bias used to nullify the CPD. The AC bias modulation is characterized by its depth $U_{\mathrm{mod}}$, angular frequency $\omega_{\mathrm{mod}}$, and phase $\Phi_{\mathrm{mod}}$.

Starting from Eq.~\eqref{eq:Fel_a}, the electrostatic force is commonly expanded into three terms featuring its DC, $\omega_{\mathrm{mod}}$, and $2\omega_{\mathrm{mod}}$ spectral components. In particular, its $\omega_{\mathrm{mod}}$ component reads \cite{Melitz2011a,Sadewasser2012a,Sadewasser2018a}:

\begin{equation}\label{eq:Fel_mod}
F_{\mathrm{el}}^{\omega_{\mathrm{mod}}}(z(t),t) = \frac{dC(z)}{dz}\left( V_{\mathrm{DC}} - V_{\mathrm{cpd}} \right)U_{\mathrm{mod}}\cos\left( \omega_{\mathrm{mod}}t + \Phi_{\mathrm{mod}} \right).
\end{equation}

\begin{figure}[htbp]
    \centering
    \includegraphics[width=\textwidth]{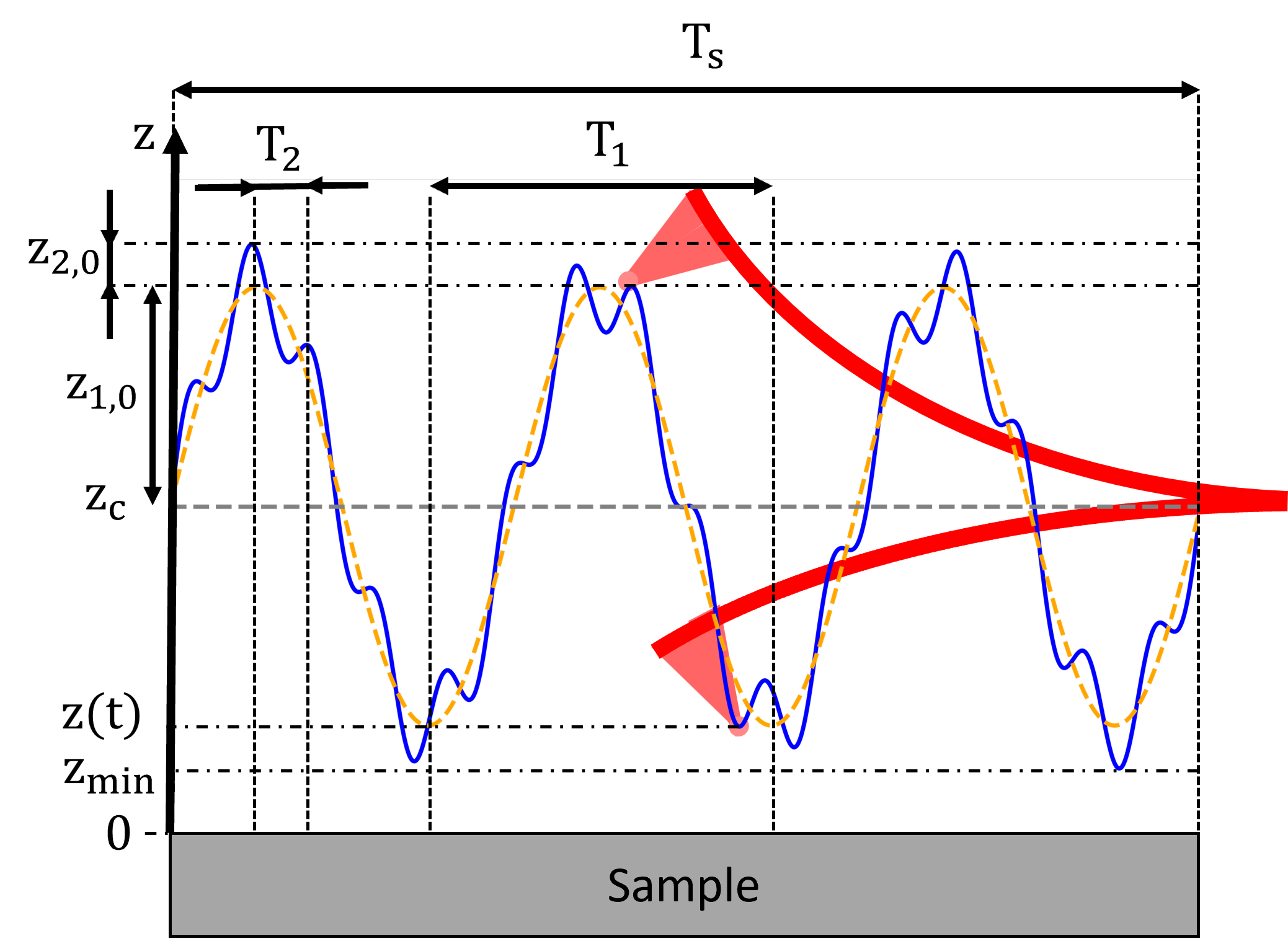}
    \caption{Geometry of the problem. Schematic representation of the cantilever oscillatory motion above the sample surface in the monomodal and bimodal regimes. The blue solid curve represents the instantaneous tip--surface distance, $z(t)$, resulting from the superposition of the first- and second-eigenmode oscillations, while the yellow dashed curve shows the first-eigenmode component. Their amplitudes are denoted $z_{1,0}$ and $z_{2,0}$, respectively. The gray dashed horizontal line indicates the average tip--surface distance, $z_c$, and $z_{\min}$ denotes the minimum tip--surface distance. The corresponding mechanical periods are $T_1=2\pi/\omega_1$ and $T_2=2\pi/\omega_2$, while $T_s$ is the super-period of the bimodal motion.}
    \label{fig:geometry_problem}
\end{figure}

A common but incomplete interpretation assumes that demodulating the $\omega_{\mathrm{mod}}$ component directly provides the KPFM controller with a signal proportional to $V_{\mathrm{DC}}-V_{\mathrm{cpd}}$. This interpretation neglects the time dependence of the CG through the instantaneous tip-surface distance, $z(t)$. The capacitance and its n\textsuperscript{th}-order spatial derivatives therefore exhibit an implicit time dependence that must be determined to establish the proper expression of the $\omega_{\mathrm{mod}}$ component of the electrostatic force. The capacitance is fundamentally a function of the tip-surface distance, and its spatial derivatives are denoted by $C^{(n)}(z)=\mathrm{d}^{n}C(z)/\mathrm{d}z^{n}$. When evaluated along the trajectory $z(t)$, they are written explicitly as $C^{(n)}(z(t))$. For compactness, the corresponding time-dependent quantities are hereafter denoted by $C^{(n)}(t,z_c)$, with $C^{(n)}(t,z_c)\equiv C^{(n)}(z(t))$, where $z_c$ denotes the average tip-surface distance of the oscillatory trajectory (see Fig.~\ref{fig:geometry_problem}).

In most situations, the cantilever dynamics is monomodal. This applies to amplitude-modulation AFM (AM-AFM) and frequency-modulation AFM (FM-AFM, i.e., non-contact AFM (nc-AFM)). The cantilever is then excited at a single angular frequency $\omega_1$, usually matching the angular resonance frequency of its first eigenmode. The instantaneous tip-surface distance is then written in terms of the average tip-surface distance $z_c$ and the minimum tip-surface distance $z_{\min}$ reached during the oscillation cycle (see Fig.~\ref{fig:geometry_problem}):

\begin{subequations}\label{eq:z}
\begin{empheq}[left=\empheqlbrace]{align}
z(t) &= z_{c} + z_{1,0}\cos\left( \omega_{1}t + \Phi_{1} \right),
\label{eq:z_a} \\
z_{c} &= z_{\min} + z_{1,0}.
\label{eq:z_b}
\end{empheq}
\end{subequations}

In Eq.~\eqref{eq:z}, the first-eigenmode displacement is characterized by its oscillation amplitude $z_{1,0}$ and phase lag $\Phi_1$ relative to the excitation waveform.

The time dependence of the CG is commonly described using a first-order truncated Taylor-series expansion (TSE) about $z_c$. Under the assumption $z_{1,0} \ll z_c$, i.e., within a ``small-oscillation-amplitude regime'', this expansion reads:

\begin{align}\label{eq:CG_Small_amplitude}
C^{(1)}(t,z_c)&\approx C^{(1)}\left( z_{c} \right) + C^{(2)}\left( z_{c} \right)\left( z(t) - z_{c} \right) \notag \\
          &\approx C^{(1)}\left( z_{c} \right) + z_{1,0}C^{(2)}\left( z_{c} \right)\cos\left( \omega_{1}t + \Phi_{1} \right).
\end{align}

Provided that this expansion is valid, Eq.~\eqref{eq:CG_Small_amplitude} describes the occurrence of lateral sidebands around the resonance frequency in the spectrum of the cantilever-displacement signal. These sidebands are induced by modulated electrostatic force components emerging when Eq.~\eqref{eq:CG_Small_amplitude} is substituted for the CG term in Eq.~\eqref{eq:Fel_mod}:

\begin{equation}
\label{eq:Fel_sidebands}
F_{\mathrm{el}}^{\omega_1\pm\omega_{\mathrm{mod}}}(z_c,t)
=
z_{1,0}\alpha_1(z_c)
\cos\left[
\left(
\omega_1\pm\omega_{\mathrm{mod}}
\right)t
+
\left(
\Phi_1\pm\Phi_{\mathrm{mod}}
\right)
\right],
\end{equation}

\noindent with:

\begin{equation}\label{eq:Fel_sidebands_alpha1}
\alpha_1(z_c)
=
\left|
\frac{C^{(2)}\left( z_{c} \right)}{2}
\left( V_{\mathrm{DC}} - V_{\mathrm{cpd}} \right)
U_{\mathrm{mod}}
\right|.
\end{equation}

The coefficient $\alpha_1(z_c)$ is thus defined as a non-negative amplitude of the electrostatic force gradient. The sign of the prefactor inside the absolute value is represented by an additional phase shift of $0$ or $\pi$. This shift is included in the effective phase of the corresponding sideband component in Eq.~\eqref{eq:Fel_sidebands}.

Demodulating the amplitude of these sidebands provides access to the CPD, while the proportionality to $\left|C^{(2)}(z_c)\right|$ makes the measurement force-gradient-sensitive.

In amplitude-modulation heterodyne KPFM (AM-He-KPFM) \cite{Sugawara2012a}, resonant detection is achieved by adjusting the bias-modulation frequency so that one of the sidebands matches the angular resonance frequency of the second eigenmode, $\omega_{2,0}$. This condition is fulfilled when $\omega_{\mathrm{mod}}=\omega_{2,0}-\omega_1$ (see Eq.~\eqref{eq:Fel_sidebands}). When the first eigenmode is also driven at resonance, $\omega_1=\omega_{1,0}$. This relation then becomes $\omega_{\mathrm{mod}}=\omega_{2,0}-\omega_{1,0}$, and the induced second-eigenmode component satisfies $\omega_2=\omega_{2,0}$. The cantilever dynamics then becomes bimodal.

The spectral description of the CG dynamics, and consequently of the electrostatic force acting on the cantilever, is therefore central to AM-He-KPFM. A first-order truncated TSE is straightforward to calculate once a capacitance model has been established. However, it may be insufficient to account for the dependence of the CG dynamics on both $z_{1,0}$ and $z_c$. This raises the question of the range of validity of the ``small-oscillation-amplitude regime'' and of the convergence of the TSE for the chosen capacitance model.

In monomodal operation, the cantilever motion is periodic at $\omega_1$. In bimodal operation, the motion is strictly periodic only when $\omega_1$ and $\omega_2$ are commensurate. Incommensurate eigenfrequencies instead produce a quasi-periodic motion without a finite super-period (see Sec.~III.A of the supplementary information (SI) file). A conventional Fourier-series expansion (FSE) over a finite period is therefore applicable to monomodal and commensurate bimodal trajectories, but not to incommensurate bimodal trajectories. For periodic trajectories, the FSE is valid regardless of the oscillation amplitudes and the average tip-surface distance $z_c$, although the dependence on these parameters is concealed in its Fourier coefficients. Conversely, the TSE explicitly preserves the dependence on the higher-order spatial derivatives of the CG evaluated at $z_c$ and remains applicable independently of the frequency relationship when considered in its non-truncated form. The two approaches thus provide complementary descriptions for periodic trajectories. The central question is then to determine the truncation order required for the TSE to faithfully reproduce the CG described by the FSE.

Despite these considerations, the CG dynamics has so far mainly been described using a first-order truncated TSE \cite{Bonnell2012a,Borgani2014a,Axt2018a,Garrett2018a,Dwyer2019a,Borgani2019a,Garrett2019a}. In these works, the concept of a ``small-oscillation-amplitude regime'' remains vaguely defined, and the convergence of the TSE has, to the best of our knowledge, never been addressed. Borgani et al. \cite{Borgani2014a} nevertheless stated that, in the case of a non-truncated TSE, the moduli of the Fourier coefficients describing the CG are linear combinations of n\textsuperscript{th}-order derivatives of the CG. Garrett et al. also demonstrated, in a related context, the importance of considering higher-order spatial derivatives of the CG \cite{Garrett2018a}.

A generalized framework for describing CG dynamics in AM-He-KPFM independently of the cantilever oscillation amplitude was therefore established. In Sec.~\ref{sec:analytical_approach}, a general non-truncated TSE of the CG is formulated (Sec.~\ref{sec:general_Taylor_description}). The realistic Hudlet-based CG model used throughout the work is then introduced (Sec.~\ref{sec:Hudlet_CG_model}). Finally, the convergence of the Taylor representation along physically admissible monomodal and bimodal trajectories is established (Sec.~\ref{sec:Taylor_series_convergence}). In the monomodal case (Sec.~\ref{sec:monomodal_CG}), the Fourier coefficients of interest are expressed analytically as combinations of $C^{(n)}(z_c)$. Term-significance criteria are then introduced to define the corresponding order-truncation regimes (OTRs). The Taylor-Fourier strategy is then extended to bimodal motion (Sec.~\ref{sec:bimodal_CG}) to derive the effective CG coefficients associated with the static, first-eigenmode, and second-eigenmode components. In Sec.~\ref{sec:numerical_results}, the convergence of the Taylor-based coefficients toward their Fourier counterparts is confirmed by numerical simulations, and the resulting OTR hierarchy is illustrated.

The bimodal extension is particularly relevant to open-loop AM-He-KPFM \cite{Grevin2023a}, in which the second eigenmode reaches a finite amplitude and contributes to the tip-surface distance modulation. The effective CG coefficients derived here provide the analytical basis required to describe the coupled nc-AFM observables, including the first-eigenmode frequency shift and dissipation and the second-eigenmode amplitude and phase. This coupling underlies the direct and inverse heterodyne effects investigated in the companion manuscript submitted concurrently \cite{PRX_A}. More generally, the present framework describes the CG dynamics of an AFM probe oscillating with one or several mechanical components. It can be extended to frequency-conversion processes, including sideband generation in amplitude-modulation heterodyne photo-induced force microscopy \cite{Jahng2016a}. The main conclusions are summarized in Sec.~\ref{sec:conclusion}, while the detailed analytical derivations and numerical developments are provided in the SI file, in particular in Secs.~I-III.H.

\clearpage

\section{Analytical approach to the capacitance gradient in monomodal and bimodal AFM}\label{sec:analytical_approach}

\subsection{General capacitance-gradient framework and Taylor-series convergence}
\label{sec:general_CG_framework}

\subsubsection{General Taylor-series description of the capacitance gradient}
\label{sec:general_Taylor_description}

Independently of the number of mechanical components involved in the cantilever motion, the instantaneous tip-surface distance can be written as:

\begin{equation}\label{eq:general_tip_surface_distance}
z(t)=z_c+\delta z(t),
\end{equation}

\noindent where $z_c$ is the average tip-surface distance and $\delta z(t)=z(t)-z_c$ denotes the complete oscillatory displacement. In the monomodal case, $\delta z(t)=z_{1,0}\cos(\omega_1t+\Phi_1)$ (see Eq.~\eqref{eq:z_a}). In the bimodal case, it is the sum of the first- and second-eigenmode displacements (see Eq.~\eqref{eq:bimodal_z}). The non-truncated TSE of the CG about $z_c$ therefore takes the general form:

\begin{equation}\label{eq:CG_Taylor}
C^{(1)}(t,z_c)
\equiv
C^{(1)}\left(z(t)\right)
=
C^{(1)}\left( z_{c} \right)
+
\sum_{n = 1}^{+ \infty}
\frac{C^{(n + 1)}\left( z_{c} \right)}{n!}
\left[ z(t) - z_{c} \right]^{n},
\end{equation}

\noindent where $n \in \mathbb{N}^{*}$ denotes the Taylor expansion order. This equation provides the common starting point for the monomodal and bimodal developments in Secs.~\ref{sec:monomodal_CG} and \ref{sec:bimodal_CG}. These developments differ through the expression of $\delta z(t)$ and the spectral components generated by the powers $[\delta z(t)]^n$.

\subsubsection{Hudlet-based capacitance-gradient model}
\label{sec:Hudlet_CG_model}

The approach described in Sec.~\ref{sec:general_Taylor_description} assumes that the TSE represents the function used to model the tip-surface capacitance within its radius of convergence. This remains a formal assumption unless the underlying function is explicitly specified. Extending the formalism to arbitrary physically admissible oscillation amplitudes therefore requires verifying that the convergence condition is fulfilled for a realistic tip-surface CG model.

\begin{figure}[htbp]
    \centering
    \includegraphics[width=\textwidth]{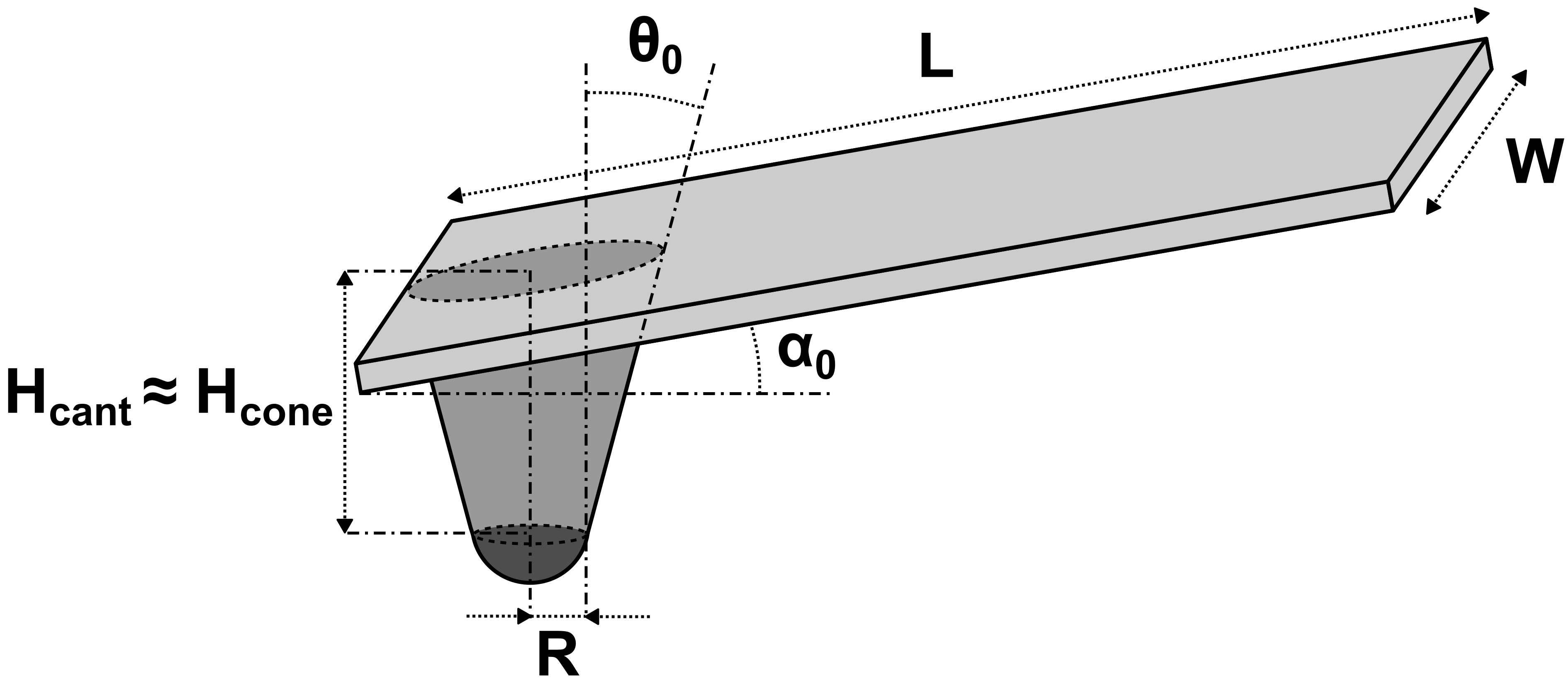}
    \caption{Tip--cone--cantilever geometry used in the Hudlet-based CG model. The schematic is not drawn to scale. The AFM probe comprises a rectangular cantilever of length $L$ and width $W$, tilted by an angle $\alpha_0$ with respect to the horizontal. The tip is described by a truncated cone of height $H_{\mathrm{cone}}$ and half-aperture angle $\theta_0$, terminated by a spherical apex of radius $R$. The lever height at the tip position, $H_{\mathrm{cant}}$, is assumed to be approximately equal to $H_{\mathrm{cone}}$.}
    \label{fig:tip_cone_cantilever_geometry}
\end{figure}

In the following, a CG model based on the work by Hudlet et al. \cite{Hudlet1998a} is used. In that work, an analytical expression was derived for the electrostatic force between a metallic tip and a metallic surface. The tip comprises a truncated conical part of height $H_{\mathrm{cone}}$ and angular aperture $2\theta_{0}$, and a spherical apex of radius $R$. The corresponding tip-cone geometry, together with the rectangular cantilever contribution included in the present model, is shown in Fig.~\ref{fig:tip_cone_cantilever_geometry}. The model is used here for positive tip-surface distances and over the physically admissible distance range sampled by the oscillating probe. Note that the electrostatic-force expression in Hudlet's original paper contained an error; its correction is detailed in Sec.~I.B of the SI file. The CG expression is further completed by a contribution from the rectangular cantilever of width $W$ and length $L$:

\begin{equation}\label{eq:CG_Hudlet}
C^{(1)}(z) = C^{(1)}_{\mathrm{cant}}(z) + C^{(1)}_{\mathrm{cone}}(z) + C^{(1)}_{\mathrm{apex}}(z),
\end{equation}

\noindent with:

\begin{subequations}\label{eq:CG_Hudlet_components}
\begin{empheq}[left=\empheqlbrace]{align}
C^{(1)}_{\mathrm{cant}}(z)
&= -\epsilon_0
\frac{WL}{(z+H_{\mathrm{cant}})(z+H_{\mathrm{cant}}+L\sin\alpha_0)},
\\
C^{(1)}_{\mathrm{cone}}(z)
&= - 2\pi \epsilon_0 g_{\theta_0} \left[
\ln\left(\frac{H_{\mathrm{cone}}}{z+a}\right) - 1
+ \frac{R\cos^2 \theta_0}{\sin \theta_0}\frac{1}{z+a}
\right],
\\
C^{(1)}_{\mathrm{apex}}(z)
&= - 2\pi \epsilon_0 R \left( \frac{1}{z} - \frac{1}{z+a} \right).
\end{empheq}
\end{subequations}

In Eqs.~\eqref{eq:CG_Hudlet_components}, $H_{\mathrm{cone}}$ is the cone height, $H_{\mathrm{cant}}$ is the lever height at the tip position, assumed here to be approximately equal to $H_{\mathrm{cone}}$, $\alpha_0$ is the angle made by the cantilever with the horizontal, $a=R(1-\sin\theta_0)$, and $g_{\theta_0}=\left[\ln\left(\tan(\theta_0/2)\right)\right]^{-2}$ is the dimensionless geometrical factor associated with the conical part of the tip.

The three CG contributions are weighted linear combinations of only two elementary functions:

\begin{equation}\label{eq:elementary_CG_functions}
h_1(z)
=
\frac{1}{z+\xi},
\qquad
h_2(z)
=
\ln(z+a),
\end{equation}

\noindent with:

\begin{equation}\label{eq:elementary_CG_xi}
\xi=0,a,H_{\mathrm{cant}},~\text{or}~H_{\mathrm{cant}}+L\sin \alpha_0.
\end{equation}

\subsubsection{Convergence of the Taylor-series expansion}
\label{sec:Taylor_series_convergence}
\label{sec:radius_of_convergence_of_the_taylor_series_expansion}

For each elementary function in Eq.~\eqref{eq:elementary_CG_functions}, the TSE convergence condition can be shown to be fulfilled along every physically admissible monomodal or bimodal trajectory. Let $z_{\mathrm{dyn}}$ denote the maximum oscillatory excursion, equal to $z_{1,0}$ in the monomodal case and to $z_{1,0}+z_{2,0}$ in the bimodal case. The physically admissible trajectory condition is:

\begin{equation}\label{eq:general_physical_trajectory_condition}
z_{\min}=z_c-z_{\mathrm{dyn}}>0.
\end{equation}

This condition ensures that the tip remains at a strictly positive distance from the surface throughout the motion.

This result is first demonstrated for the rational function $h_1(z)$. Introducing the change of variable:

\begin{equation}\label{eq:general_convergence_x}
x=\frac{\delta z(t)}{z_c},
\end{equation}

\noindent and defining:

\begin{equation}\label{eq:general_convergence_beta}
\beta=1+\frac{\xi}{z_c},
\end{equation}

\noindent one obtains:

\begin{equation}\label{eq:general_h1_x}
h_1(z)
=
\frac{1}{z+\xi}
\rightarrow
h_1(x)
=
\frac{1}{z_c}
\frac{1}{\beta+x}.
\end{equation}

The radius of convergence is $\beta$ because the nearest singularity to the expansion center $x=0$ is located at $x=-\beta$. Moreover,

\begin{equation}\label{eq:general_convergence_bound}
|x|
=\left|\frac{\delta z(t)}{z_c}\right|
\leq\frac{z_{\mathrm{dyn}}}{z_c}
=\frac{z_{\mathrm{dyn}}}{z_{\min}+z_{\mathrm{dyn}}}
<1\leq\beta.
\end{equation}

Thus, the TSE provides an exact representation of $h_1(z(t))$ over the complete trajectory and leads to Eq.~\eqref{eq:h1} (see Sec.~I.D.1 of the SI file):

\begin{equation}\label{eq:h1}
h_1(z(t))
=
h_1(z_c)
+
\sum_{n=1}^{+\infty}
\frac{h_1^{(n)}(z_c)}{n!}
[\delta z(t)]^n.
\end{equation}

In the monomodal case, $\delta z(t)=z_{1,0}\cos(\omega_1t+\Phi_1)$, so Eq.~\eqref{eq:h1} becomes:

\begin{equation}\label{eq:h1_monomodal}
h_1(z(t))
=
h_1(z_c)
+
\sum_{n=1}^{+\infty}
\frac{z_{1,0}^{n}}{n!}
h_1^{(n)}(z_c)
\cos^{n}\left(\omega_1t+\Phi_1\right).
\end{equation}

This is the form used in the monomodal Taylor-Fourier development.

The same approach shows that the logarithmic function $h_2(z)$ is exactly represented by a non-truncated TSE. Indeed, with $x=\delta z(t)/(z_c+a)$, the physical condition in Eq.~\eqref{eq:general_physical_trajectory_condition} gives:

\begin{equation}\label{eq:general_logarithmic_convergence_bound}
|x|
\leq
\frac{z_{\mathrm{dyn}}}{z_c+a}
=
\frac{z_{\mathrm{dyn}}}{z_{\min}+z_{\mathrm{dyn}}+a}
<1.
\end{equation}

The detailed derivation is provided in Sec.~I.D.2 of the SI file. Since the CG is a linear combination of these elementary functions, the general expansion in Eq.~\eqref{eq:CG_Taylor} is consistently recovered. Its convergence is therefore ensured for both monomodal and bimodal motion, regardless of the oscillation amplitudes, provided that the complete trajectory remains physically admissible, i.e. $z_{\min}>0$.

\subsection{Capacitance-gradient dynamics in monomodal AFM}
\label{sec:monomodal_CG}

\subsubsection{Taylor-Fourier correspondence and effective capacitance-gradient coefficients}
\label{sec:monomodal_Taylor_Fourier}

In monomodal AFM, the tip-surface CG is coupled to the periodic mechanical oscillation of the cantilever at the angular frequency $\omega_1$ (see Eq.~\eqref{eq:z}). It can be described by a FSE, valid for any oscillation amplitude $z_{1,0}$:

\begin{equation}\label{eq:CG_Fourier}
C^{(1)}(t,z_c) = X_{0} + \sum_{m = 1}^{+ \infty} X_{m}\cos\left( m\omega_{1}t + \varphi_{m} \right),
\end{equation}

\noindent where $X_0$ is the static Fourier coefficient of the CG, while $X_m$ and $\varphi_m$ denote the signed Fourier coefficient and associated phase of its $m$-th harmonic component ($m \in \mathbb{N}^*$), respectively. The magnitude of $X_m$ gives the amplitude of the corresponding component, whereas a negative sign can equivalently be absorbed into $\varphi_m$ through an additional phase shift of $\pi$. Thus, $X_m$ should not be interpreted as a strictly non-negative amplitude. These quantities can be determined by specializing the general non-truncated TSE introduced in Eq.~\eqref{eq:CG_Taylor} to the monomodal trajectory. Substituting the monomodal expression of $z(t)$ (see Eq.~\eqref{eq:z_a}) into the general expansion gives:

\begin{equation}\label{eq:CG_z_nth_power}
C^{(1)}(t,z_c) = C^{(1)}\left( z_{c} \right) + \sum_{n = 1}^{+ \infty}\frac{C^{(n + 1)}\left( z_{c} \right)}{n!}z_{1,0}^{n}\cos^{n}\left( \omega_{1}t + \Phi_{1} \right).
\end{equation}

As pointed out by Borgani et al. \cite{Borgani2014a}, Eq.~\eqref{eq:CG_z_nth_power} suggests that the coefficients $X_m$ are linear combinations of derivative-amplitude products. These products involve n\textsuperscript{th}-order spatial derivatives of the CG and powers of order $n$ of the oscillation amplitude.

The most relevant CG components in monomodal AFM are the static component and the fundamental component at $\omega_1$. Their calculation from Eq.~\eqref{eq:CG_z_nth_power}, based on Newton's binomial formula, is detailed in Secs.~II.B-II.D of the SI file. Retaining only these two spectral components gives the compact representation:

\begin{equation}\label{eq:CG_Taylor_DC_1st_harmonic}
C^{(1)}(t,z_c)
\approx
K_{0}(z_{c})
+
z_{1,0}K_{1}(z_{c})
\cos\left(\omega_{1}t + \Phi_{1}\right).
\end{equation}

Here, the symbol $\approx$ denotes a spectral truncation restricted to the static and fundamental components. It does not correspond to a small-oscillation-amplitude approximation: the higher harmonics at $m\omega_1$, with $m\geq2$, are omitted from this compact representation, whereas the coefficients $K_0(z_c)$ and $K_1(z_c)$ are derived exactly from the non-truncated TSE. The coefficients $K_i(z_c)$ ($i=0,1$) are the effective CG coefficients associated with the static and $\omega_1$ components, respectively. Their expressions are:

\begin{subequations}\label{eq:CG_Taylor_Coefs}
\begin{empheq}[left=\empheqlbrace]{align}
K_{0}\left( z_{c} \right) &= C^{(1)}\left( z_{c} \right) + \Lambda_{0}\left( z_{c} \right),
\label{eq:CG_Taylor_Coefs_K0} \\
K_{1}\left( z_{c} \right) &= C^{(2)}\left( z_{c} \right) + \Lambda_{1}\left( z_{c} \right),
\label{eq:CG_Taylor_Coefs_K1}
\end{empheq}
\end{subequations}

\noindent where $\Lambda_i(z_c)$ ($i=0,1$) collect the higher-order correction terms:

\begin{subequations}\label{eq:CG_Taylor_Lambda}
\begin{empheq}[left=\empheqlbrace]{align}
\Lambda_{0}\left( z_{c} \right) &= \sum_{q = 1}^{+ \infty}\frac{z_{1,0}^{2q}}{{2^{2q}(q!)}^{2}}C^{(2q + 1)}\left( z_{c} \right),
\label{eq:CG_Taylor_Lambda0} \\
\Lambda_{1}\left( z_{c} \right) &= \sum_{q = 1}^{+ \infty}\frac{z_{1,0}^{2q}}{2^{2q}q!(q + 1)!}C^{(2q + 2)}\left( z_{c} \right).
\label{eq:CG_Taylor_Lambda1}
\end{empheq}
\end{subequations}

Their detailed derivation is provided in Sec.~II.D.2 of the SI file.

Comparing the static and fundamental components of Eq.~\eqref{eq:CG_Fourier} with those retained in Eq.~\eqref{eq:CG_Taylor_DC_1st_harmonic} gives the first two Fourier coefficients of the CG:

\begin{subequations}\label{eq:monomodal_TSE_FSE}
\begin{empheq}[left=\empheqlbrace]{align}
X_0 &= K_0(z_c), \\
X_1 &= K_1(z_c)z_{1,0}.
\end{empheq}
\end{subequations}

More generally, the complete Fourier spectrum can be obtained by expanding the cosine powers in Eq.~\eqref{eq:CG_z_nth_power}. Even Taylor powers generate the static and even-harmonic components, whereas odd Taylor powers generate the odd-harmonic components. The derivation and general expressions of $X_0$, $X_{2m}$, and $X_{2m+1}$ are provided in Sec.~II.C of the SI file. These terms correspond to the static, even-harmonic, and odd-harmonic Fourier coefficients, respectively. The same section also gives the corresponding harmonic-selection rules and phase relations. In the following, only the static and fundamental components are required, whose correspondence with the effective Taylor-based coefficients is specified by Eq.~\eqref{eq:monomodal_TSE_FSE}. The phase is defined modulo $2\pi$.

Eqs.~\eqref{eq:CG_Taylor_DC_1st_harmonic}-\eqref{eq:CG_Taylor_Lambda} provide exact expressions for the static and fundamental CG components relevant to monomodal AM-He-KPFM, regardless of the cantilever oscillation amplitude. Consequently, the coefficient $\alpha_1(z_c)$ entering the lateral-sideband expression in Eq.~\eqref{eq:Fel_sidebands} must be replaced by:

\begin{align}\label{eq:alpha1_K1_effective}
\alpha_1(z_c)
&=
\left|
\frac{K_1(z_c)}{2}
\left( V_{\mathrm{DC}} - V_{\mathrm{cpd}} \right)
U_{\mathrm{mod}}
\right|
\notag\\
&=
\left|
\frac{C^{(2)}(z_c)+\Lambda_1(z_c)}{2}
\left( V_{\mathrm{DC}} - V_{\mathrm{cpd}} \right)
U_{\mathrm{mod}}
\right|.
\end{align}

\subsubsection{Order-truncation regimes in the monomodal case}
\label{sec:monomodal_OTR}

Reducing Eq.~\eqref{eq:CG_Taylor_DC_1st_harmonic} to the usual ``small-oscillation-amplitude regime'' expression in Eq.~\eqref{eq:CG_Small_amplitude} requires neglecting higher-order corrections. These corrections enter $K_0(z_c)$ and $K_1(z_c)$. To replace this qualitative approximation with an explicit analytical criterion, OTRs are used and assigned independently to the static and $\omega_1$ components of the CG dynamics. The OTR order refers to the truncation level of the reorganized correction series and should not be confused with the truncation order $n$ of the original TSE. In the monomodal case, a grouped level $q$ originates from the Taylor order $n=2q$ for $K_0(z_c)$ and from $n=2q+1$ for $K_1(z_c)$. The complete correspondence, including the bimodal coefficients, is summarized in Table~\ref{tab:OTR_Taylor_order_correspondence}.

For each component, the effective coefficient is written as a series of contributions grouped by correction order:

\begin{equation}
K_i(z_c)
=
\sum_{q=0}^{+\infty}
T_{i,q}(z_c),
\end{equation}

\noindent where $T_{i,0}$ is the leading term. The relative significance of the contribution of order $q$ is defined as:

\begin{equation}
r_{i,q}
=
\left|
\frac{T_{i,q}(z_c)}
{T_{i,0}(z_c)}
\right|.
\end{equation}

For a prescribed threshold $\tau$, the retained OTR order is the highest correction order whose relative contribution remains significant:

\begin{equation}
\ell_i^\star
=
\max
\left\{
q\in\mathbb{N}
\;\big|\;
r_{i,q}\geq\tau
\right\}.
\label{eq:monomodal_OTR_order}
\end{equation}

This definition assigns the zeroth-order truncation regime (ZOTR) when $\ell_i^\star=0$. It assigns the first-order truncation regime (FOTR) when $\ell_i^\star=1$. For $\ell_i^\star\geq2$, it assigns a higher-order truncation regime of order $\ell_i^\star$ (HOTR-$\ell_i^\star$).

In the ZOTR, only the leading terms are retained:

\begin{subequations}\label{eq:ZOTR_K}
\begin{empheq}[left=\empheqlbrace]{align}
K_0(z_c)
&\approx
K_0^{\mathrm{ZOTR}}(z_c)
=
C^{(1)}(z_c),
\label{eq:ZOTR_K0}
\\
K_1(z_c)
&\approx
K_1^{\mathrm{ZOTR}}(z_c)
=
C^{(2)}(z_c).
\label{eq:ZOTR_K1}
\end{empheq}
\end{subequations}

The approximation for $K_1$ corresponds directly to the conventional KPFM ``small-oscillation-amplitude regime''.

When the first correction level is significant, the FOTR retains the leading term and the first higher-order correction:

\begin{subequations}\label{eq:FOTR_K}
\begin{empheq}[left=\empheqlbrace]{align}
K_0(z_c)
&\approx
K_0^{\mathrm{FOTR}}(z_c)
=
C^{(1)}(z_c)
+
\frac{z_{1,0}^{2}}{4}
C^{(3)}(z_c),
\label{eq:FOTR_K0}
\\
K_1(z_c)
&\approx
K_1^{\mathrm{FOTR}}(z_c)
=
C^{(2)}(z_c)
+
\frac{z_{1,0}^{2}}{8}
C^{(4)}(z_c).
\label{eq:FOTR_K1}
\end{empheq}
\end{subequations}

More generally, a HOTR-$\ell$ description retains all significant contributions up to correction order $\ell$. The explicit significance ratios and complete ZOTR, FOTR, and HOTR-$\ell$ attribution conditions are provided in Sec.~II.E of the SI file.

The OTR attribution depends on the oscillation amplitude $z_{1,0}$, the average tip-surface distance $z_c$, the prescribed threshold $\tau$, and the selected CG model through its successive derivatives. Increasing the oscillation amplitude and/or decreasing the average tip-surface distance generally enhances the weight of higher-order terms because the cantilever samples a broader and more nonlinear portion of the CG curve. The dynamics consequently becomes more sensitive to the short-range electrostatic contributions contained in the higher-order spatial derivatives of the CG. This effect is particularly important for the $\omega_1$ component because the heterodyne sidebands are governed by $K_1(z_c)$ and its correction term $\Lambda_1(z_c)$.

The KPFM meaning of ``small-oscillation-amplitude regime'' should not be confused with its usual meaning in nc-AFM. In nc-AFM, this regime refers to amplitudes comparable to or smaller than the decay length of short-range tip-surface interactions, typically $z_{1,0}\leq2~\text{\AA}$. In the present context, it refers instead to the hierarchy of the CG-derivative contributions entering the effective coefficients and therefore depends on the complete probe geometry and capacitance model.

\subsection{Capacitance-gradient dynamics in bimodal AFM}
\label{sec:bimodal_CG}

\subsubsection{Foundations and spectral organization of bimodal AFM}
\label{sec:bimodal_foundations}

To the best of our knowledge, CG dynamics has not yet been examined within the framework of bimodal AFM theory, whose foundations are laid in Refs.~\cite{Rodriguez2004a,Lozano2008a,Lozano2009a}.

In bimodal AFM, the cantilever is simultaneously driven in two eigenmodes. The first eigenmode is used to track the sample topography through an active $z$-feedback loop in either AM-AFM or nc-AFM. The physical observables of the second eigenmode, namely its oscillation amplitude and phase or its resonance frequency, are then used to probe tip-surface interactions without being hindered by the topographic feedback. Open-loop AM-He-KPFM is a particular bimodal configuration in which the second eigenmode is excited through an electrostatic heterodyne coupling effect \cite{PRX_A}.

The oscillatory cantilever displacement is approximated as the superposition of the oscillations induced by each eigenmode. The instantaneous tip-surface distance $z(t)$ is therefore defined as:

\begin{subequations}\label{eq:bimodal_z}
\begin{empheq}[left=\empheqlbrace]{align}
z(t) &= z_{c} + \sum_{i=1}^{2} z_i(t),
\label{eq:bimodal_z(t)} \\
z_{c} &= z_{\min} + \sum_{i=1}^{2} z_{i,0},
\label{eq:bimodal_zc} \\
z_{i}(t) &= z_{i,0}\cos\left( \omega_{i}t + \Phi_{i} \right).
\label{eq:bimodal_zi}
\end{empheq}
\end{subequations}

Each eigenmode oscillation is characterized by its amplitude, angular frequency, and phase, denoted $z_{i,0}$, $\omega_i$, and $\Phi_i$, respectively. In the numerical illustrations, a rectangular cantilever whose first two angular resonance frequencies satisfy $\omega_{2,0}=6.3\omega_{1,0}$ is used. The prescribed mechanical components are taken at resonance, such that $\omega_1=\omega_{1,0}$ and $\omega_2=\omega_{2,0}$. The analytical framework is not restricted to this ratio and applies to integer-multiple, commensurate, and incommensurate frequency relationships. The influence of frequency commensurability is discussed in detail in Sec.~III.A of the SI file.

For commensurate eigenfrequencies, two coprime positive integers $p_1$ and $p_2$ exist such that $\omega_1=p_1\omega_s$ and $\omega_2=p_2\omega_s$, where $\omega_s$ is the angular super-frequency. The cantilever dynamics is then characterized by the super-period $T_s$ \cite{Lozano2009a,Kawai2009a,Lai2015a}:

\begin{equation}\label{eq:bimodal_Superperiod}
T_s
=
\frac{2\pi}{\omega_s}
=
p_1T_1
=
\frac{2\pi p_1}{\omega_1}
=
p_2T_2
=
\frac{2\pi p_2}{\omega_2},
\qquad
p_1,p_2\in\mathbb{N}^{\ast},
\qquad
\gcd\left(p_1,p_2\right)=1.
\end{equation}

The mechanical motion, and consequently the CG dynamics, is then $T_s$-periodic, with prominent spectral components at $\omega_1$ and $\omega_2$. Owing to the time dependence of the CG, these components generate sidebands at $(\omega_1\pm\omega_{\mathrm{mod}})$ and $(\omega_2\pm\omega_{\mathrm{mod}})$. This follows from Eq.~\eqref{eq:Fel_sidebands} \cite{PRX_A}.

\subsubsection{Taylor-Fourier correspondence and effective capacitance-gradient coefficients}
\label{sec:bimodal_Taylor_Fourier}

For commensurate eigenfrequencies, the CG is $T_s$-periodic and can be expanded as a Fourier series:

\begin{equation}\label{eq:bimodal_CG_Fourier}
C^{(1)}(t,z_c) = X_{0} + \sum_{m = 1}^{+ \infty} X_{m}\cos\left( m\omega_{s}t + \varphi_{m} \right).
\end{equation}

The signed-coefficient convention introduced in the monomodal case is retained: $X_m$ is not constrained to be positive, and its sign can equivalently be absorbed into $\varphi_m$ through an additional phase shift of $\pi$. Within this periodic case, Eq.~\eqref{eq:bimodal_CG_Fourier} is valid for arbitrary eigenmode oscillation amplitudes. Using Eq.~\eqref{eq:bimodal_Superperiod}, it can be rewritten as:

\begin{align}
C^{(1)}(t,z_c)
&=
X_{0}
+
X_{p_{1}}\cos\left(p_{1}\omega_{s}t+\varphi_{p_{1}}\right)
+
X_{p_{2}}\cos\left(p_{2}\omega_{s}t+\varphi_{p_{2}}\right)
\notag \\
&\quad
+
\sum_{\substack{m=1 \\ m\ne p_{1},p_{2}}}^{+\infty}
X_{m}\cos\left(m\omega_{s}t+\varphi_{m}\right)
\notag \\
&=
X_{0}
+
X_{p_{1}}\cos\left(\omega_{1}t+\varphi_{p_{1}}\right)
+
X_{p_{2}}\cos\left(\omega_{2}t+\varphi_{p_{2}}\right)
\notag \\
&\quad
+
\sum_{\substack{m=1 \\ m\ne p_{1},p_{2}}}^{+\infty}
X_{m}\cos\left(m\omega_{s}t+\varphi_{m}\right).
\label{eq:bimodal_CG_Fourier_prominent_components}
\end{align}

Following the monomodal approach, the Fourier coefficients are derived from the general non-truncated TSE in Eq.~\eqref{eq:CG_Taylor}. Substituting the bimodal cantilever displacement from Eq.~\eqref{eq:bimodal_z} gives:

\begin{equation}\label{eq:bimodal_CG_Taylor}
C^{(1)}\left(t,z_c\right)
=
C^{(1)}\left(z_{c}\right)
+
\sum_{n = 1}^{+\infty}
\frac{C^{(n + 1)}\left(z_{c}\right)}{n!}
{\left[
z_{1,0}\cos\left(\omega_{1}t+\Phi_{1}\right)
+
z_{2,0}\cos\left(\omega_{2}t+\Phi_{2}\right)
\right]}^{n}.
\end{equation}

Its convergence along the complete bimodal trajectory follows from the general result established in Sec.~\ref{sec:Taylor_series_convergence}. The same master equation is obtained by expanding the elementary functions $h_i(z)$ within their convergence radii; the detailed derivation is provided in Sec.~III.B.1 of the SI file.

Products of cosines generate spectral components at integer combinations of the eigenmode angular frequencies, $n_1\omega_1+n_2\omega_2$, with $(n_1,n_2)\in\mathbb{Z}^2$. Each contribution carries a phase of the form $n_1\Phi_1+n_2\Phi_2$, up to a constant shift introduced by the cosine-product identities. The mixed terms therefore produce an ordered set of spectral phases dictated by integer combinations of $\Phi_1$ and $\Phi_2$ and by the expansion order $n$. This organization is examined numerically in Sec.~\ref{sec:spectral_content_bimodal}.

The static, $\omega_1$, and $\omega_2$ components of the CG are retained. Their detailed derivation from the frequency-selection rules is provided in Secs.~III.C-III.E of the SI file. Restricting the spectral representation to these three components gives:

\begin{align}\label{eq:bimodal_Taylor_prominent_components}
C^{(1)}\left(t,z_c\right)
&\approx
K_{0}(z_{c})
+
K_{1}\left(z_{c}\right)z_{1}(t)
+
K_{2}\left(z_{c}\right)z_{2}(t)
\notag \\
&=
K_{0}(z_{c})
+
z_{1,0}K_{1}(z_{c})
\cos\left(\omega_{1}t+\Phi_{1}\right)
+
z_{2,0}K_{2}(z_{c})
\cos\left(\omega_{2}t+\Phi_{2}\right).
\end{align}

Here, the symbol $\approx$ denotes a spectral truncation restricted to the static, $\omega_1$, and $\omega_2$ components. The higher harmonics and intermodulation components at the other integer combinations of $\omega_1$ and $\omega_2$ are omitted from this compact representation. This notation does not imply a small-oscillation-amplitude or Taylor-order approximation. The coefficients $K_i(z_c)$ $(i=0,1,2)$ are the effective coefficients associated with the retained components and depend on whether $\omega_1$ and $\omega_2$ are integer-multiple, commensurate, or incommensurate, as detailed in Secs.~III.C-III.D of the SI file. In compact form, they read:

\begin{subequations}\label{eq:bimodal_Taylor_Coefs}
\begin{empheq}[left=\empheqlbrace]{align}
K_{0}\left(z_{c}\right)
&=
C^{(1)}\left(z_{c}\right)
+
\Lambda_{0}\left(z_{c}\right),
\label{eq:bimodal_Taylor_Coefs_K0} \\
K_{1}\left(z_{c}\right)
&=
C^{(2)}\left(z_{c}\right)
+
\Lambda_{1}\left(z_{c}\right),
\label{eq:bimodal_Taylor_Coefs_K1} \\
K_{2}\left(z_{c}\right)
&=
C^{(2)}\left(z_{c}\right)
+
\Lambda_{2}\left(z_{c}\right).
\label{eq:bimodal_Taylor_Coefs_K2}
\end{empheq}
\end{subequations}

The coefficients $\Lambda_i(z_c)$ $(i=0,1,2)$ contain higher-order spatial derivatives of the CG weighted by products of the eigenmode amplitudes. The compact expressions derived for incommensurate frequencies are used here. Their derivation is detailed in Sec.~III.E of the SI file, while the influence of commensurability-induced contributions and their absence over the Taylor-order range considered here are discussed in Sec.~III.G.2 of the SI file. For the frequency ratio and Taylor-order range considered here, these additional contributions do not enter the coefficients of interest:

\begin{subequations}\label{eq:bimodal_Lambda}
\begin{empheq}[left=\empheqlbrace]{align}
\Lambda_{0}\left(z_{c}\right)
&=
\sum_{m = 1}^{+\infty}
\sum_{q = 0}^{m}
\frac{C^{(2m + 1)}\left(z_{c}\right)}{2^{2m}}
\frac{z_{1,0}^{2q}z_{2,0}^{2(m-q)}}
{\left[q!(m-q)!\right]^{2}},
\label{eq:bimodal_Lambda0} \\
\Lambda_{1}\left(z_{c}\right)
&=
\sum_{m = 1}^{+\infty}
\sum_{q = 0}^{m}
\frac{C^{(2m + 2)}\left(z_{c}\right)}{2^{2m}}
\frac{z_{1,0}^{2q}z_{2,0}^{2(m-q)}}
{q!(q+1)!\left[(m-q)!\right]^{2}},
\label{eq:bimodal_Lambda1} \\
\Lambda_{2}\left(z_{c}\right)
&=
\sum_{m = 1}^{+\infty}
\sum_{q = 0}^{m}
\frac{C^{(2m + 2)}\left(z_{c}\right)}{2^{2m}}
\frac{z_{1,0}^{2q}z_{2,0}^{2(m-q)}}
{(q!)^{2}(m-q)!(m-q+1)!}.
\label{eq:bimodal_Lambda2}
\end{empheq}
\end{subequations}

The three Fourier coefficients of interest are obtained by comparing corresponding components in Eqs.~\eqref{eq:bimodal_Taylor_prominent_components} and \eqref{eq:bimodal_CG_Fourier_prominent_components}. The retained components are the static, $\omega_1$, and $\omega_2$ terms:

\begin{subequations}\label{eq:bimodal_TSE_FSE}
\begin{empheq}[left=\empheqlbrace]{align}
X_0 &= K_0(z_c), \\
X_{p_1} &= K_1(z_c)z_{1,0}, \\
X_{p_2} &= K_2(z_c)z_{2,0}.
\end{empheq}
\end{subequations}

These effective coefficients also determine the amplitudes of the electrostatic-force components obtained by combining the CG with the electrically modulated bias in Eq.~\eqref{eq:Fel}. In particular, $K_1(z_c)$ and $K_2(z_c)$ enter the coupling coefficients governing the first- and second-eigenmode dynamics. They consequently appear in the coupled nc-AFM observables measured in open-loop AM-He-KPFM: the first-eigenmode frequency shift and dissipation and the second-eigenmode amplitude and phase. The corresponding force components, observables, and direct and inverse heterodyne effects are derived and experimentally investigated in the companion manuscript \cite{PRX_A}. The coefficient $K_0(z_c)$ specifies the static CG component and the associated electrostatic-force channels.

The monomodal case is recovered by setting $z_{2,0}=0$.

\subsubsection{Order-truncation regimes in the bimodal case}
\label{sec:bimodal_OTR}

The OTR nomenclature introduced in the monomodal case extends to the bimodal coefficients $K_i(z_c)$, with $i=0,1,2$ referring to the static, $\omega_1$, and $\omega_2$ components, respectively. The higher-order corrections depend jointly on $z_{1,0}$, $z_{2,0}$, and $z_c$, as well as on the selected capacitance model and probe geometry through the successive derivatives $C^{(n)}(z_c)$. The OTR must therefore be assigned separately for each component and set of mechanical and geometrical parameters.

The terms entering each $K_i$ series are grouped according to the correction index $m$ in Eq.~\eqref{eq:bimodal_Lambda}. The bimodal OTR criterion follows from Eq.~\eqref{eq:monomodal_OTR_order} by replacing $q$ with $m$ and $T_{i,q}$ with $T_{i,m}$. ZOTR retains only the leading contribution, FOTR includes the first grouped correction, and HOTR-$\ell$ retains all significant grouped corrections up to order $\ell$. The correspondence between $m$ and the Taylor truncation order $n$ is summarized in Table~\ref{tab:OTR_Taylor_order_correspondence}. The index $m$ originates from $n=2m$ for $K_0(z_c)$. It originates from $n=2m+1$ for $K_1(z_c)$ and $K_2(z_c)$.

In the ZOTR, the effective coefficients reduce to:

\begin{subequations}\label{eq:bimodal_ZOTR_K}
\begin{empheq}[left=\empheqlbrace]{align}
K_0(z_c)
&\approx
K_0^{\mathrm{ZOTR}}(z_c)
=
C^{(1)}(z_c),
\label{eq:bimodal_ZOTR_K0}
\\
K_1(z_c)
&\approx
K_1^{\mathrm{ZOTR}}(z_c)
=
C^{(2)}(z_c),
\label{eq:bimodal_ZOTR_K1}
\\
K_2(z_c)
&\approx
K_2^{\mathrm{ZOTR}}(z_c)
=
C^{(2)}(z_c).
\label{eq:bimodal_ZOTR_K2}
\end{empheq}
\end{subequations}

When the first grouped correction is significant, the FOTR retains the leading contribution and the first correction level:

\begin{subequations}\label{eq:bimodal_FOTR_K}
\begin{empheq}[left=\empheqlbrace]{align}
K_0(z_c)
&\approx
K_0^{\mathrm{FOTR}}(z_c)
=
C^{(1)}(z_c)
+
\frac{z_{1,0}^{2}+z_{2,0}^{2}}{4}
C^{(3)}(z_c),
\label{eq:bimodal_FOTR_K0}
\\
K_1(z_c)
&\approx
K_1^{\mathrm{FOTR}}(z_c)
=
C^{(2)}(z_c)
+
\left(
\frac{z_{1,0}^{2}}{8}
+
\frac{z_{2,0}^{2}}{4}
\right)
C^{(4)}(z_c),
\label{eq:bimodal_FOTR_K1}
\\
K_2(z_c)
&\approx
K_2^{\mathrm{FOTR}}(z_c)
=
C^{(2)}(z_c)
+
\left(
\frac{z_{1,0}^{2}}{4}
+
\frac{z_{2,0}^{2}}{8}
\right)
C^{(4)}(z_c).
\label{eq:bimodal_FOTR_K2}
\end{empheq}
\end{subequations}

More generally, a HOTR-$\ell$ description retains all significant grouped contributions up to order $\ell$. The explicit grouped terms, significance ratios, and complete ZOTR, FOTR, and HOTR-$\ell$ attribution conditions are provided in Sec.~III.F of the SI file.

This definition does not require the assumption that $z_{1,0}\gg z_{2,0}$. However, this limit is relevant to many open-loop AM-He-KPFM configurations, in which the first eigenmode carries most of the mechanical amplitude while the second eigenmode exhibits a smaller but finite heterodyne response. In this limit, Eq.~\eqref{eq:bimodal_FOTR_K} shows that the first correction to $K_2(z_c)$ is twice that of $K_1(z_c)$:

\begin{equation}
\frac{z_{1,0}^{2}}{4}C^{(4)}(z_c)
=
2
\left[
\frac{z_{1,0}^{2}}{8}C^{(4)}(z_c)
\right].
\label{eq:bimodal_FOTR_correction_ratio}
\end{equation}

Consequently, the $\omega_2$ component may enter a higher OTR before the $\omega_1$ component, even when the second-eigenmode amplitude remains small. Second-eigenmode observables in open-loop AM-He-KPFM are therefore particularly sensitive to the nonlinear distance dependence of the CG and to the higher-order derivatives associated with short-range electrostatic contributions.

These analytical results are compared with numerical simulations in the next section.

\clearpage

\section{Numerical simulations}
\label{sec:numerical_results}
\label{sec:Numerical_investigations}

Closed-form expressions for the CG coefficients associated with the dominant spectral components of the cantilever dynamics are provided by the analytical framework developed in Sec.~\ref{sec:analytical_approach}. These results are now tested numerically using a realistic tip-surface interaction model and parameters representative of the nc-AFM setup operated under UHV at room temperature. Unless otherwise stated, the first- and second-eigenmode resonance frequencies are set to $f_{1,0}=150~\mathrm{kHz}$ and $f_{2,0}=6.3f_{1,0}=945~\mathrm{kHz}$, respectively. The corresponding stiffnesses are $k_1=48~\mathrm{N/m}$ and $k_2=39.3k_1\approx1886~\mathrm{N/m}$. The prescribed mechanical components are taken at resonance, such that $f_1=f_{1,0}$ and $f_2=f_{2,0}$. The average tip-surface distance is fixed at $z_c=1.6~\mathrm{nm}$ and the first-eigenmode amplitude at $z_{1,0}=1~\mathrm{nm}$. The second-eigenmode amplitude is set to $z_{2,0}=0$ in the monomodal case and to $z_{2,0}=0.1~\mathrm{nm}$ in the bimodal case. The remaining geometrical, electrostatic, and sampling parameters are introduced when required and summarized in Table~\ref{tab:numerical_parameters}.

In the analytical developments, frequencies are expressed as angular frequencies. Cyclic frequencies are used in the numerical discussion and figures. The corresponding definitions are $f_i=\omega_i/(2\pi)$ and $f_{i,0}=\omega_{i,0}/(2\pi)$ $(i=1,2)$. The set of definitions also includes $f_s=\omega_s/(2\pi)$ and $f_{\mathrm{mod}}=\omega_{\mathrm{mod}}/(2\pi)$. The quantities $f_1$ and $f_2$ denote the frequencies of the first- and second-eigenmode mechanical components, respectively, while $f_{1,0}$ and $f_{2,0}$ denote the corresponding resonance frequencies. The quantities $f_s$ and $f_{\mathrm{mod}}$ are the super-frequency associated with the bimodal super-period and the bias-modulation frequency, respectively.

In this section, the force landscape sampled by the oscillating probe is first defined, and the time-domain and spectral response of the CG in the monomodal and bimodal regimes are then analyzed. The convergence of the Taylor-based coefficients toward the Fourier coefficients extracted from the exact numerical signal is subsequently verified. The OTRs of the effective CG coefficients are then determined throughout the $(z_c,z_{1,0})$ parameter space using the analytical term-significance criterion.

\subsection{Tip-surface interaction force}
\label{sec:tip_surface_interaction_force}

The numerical tip-surface interaction landscape in which the AFM probe oscillates is first introduced. This establishes the distance range explored during the motion and identifies the force contributions acting on the cantilever. It also provides the physical context in which the distance-dependent CG is evaluated along the prescribed monomodal and bimodal trajectories.

\subsubsection{Numerical force model}

The tip-surface interaction force used in the numerical simulations is written as the sum of a long-range van der Waals contribution, a short-range Morse-like contribution, and an electrostatic contribution:

\begin{equation}\label{eq:numerical_force_ts}
F_{\mathrm{ts}}(z,t)
=
F_{\mathrm{LR}}(z)
+
F_{\mathrm{SR}}(z)
+
F_{\mathrm{el}}(z,t).
\end{equation}

The long-range contribution is described by the non-retarded sphere-plane van der Waals expression, whereas the short-range contribution is represented by an effective Morse-like force law:

\begin{subequations}\label{eq:numerical_force_LR_SR}
\begin{empheq}[left=\empheqlbrace]{align}
F_{\mathrm{LR}}(z)
&=
-\frac{H_{\mathrm{A}}R}{6z^2},
\label{eq:numerical_force_LR}
\\
F_{\mathrm{SR}}(z)
&=
-2U_0\kappa_{\mathrm{SR}}
\left[
\exp\left(-\kappa_{\mathrm{SR}}\left(z-z_{\mathrm{eq}}\right)\right)
-
\exp\left(-2\kappa_{\mathrm{SR}}\left(z-z_{\mathrm{eq}}\right)\right)
\right].
\label{eq:numerical_force_SR}
\end{empheq}
\end{subequations}

In Eqs.~\eqref{eq:numerical_force_LR_SR}, $R$ is the tip-apex radius, $H_{\mathrm{A}}$ is the Hamaker constant, $U_0$ is the Morse potential depth, $\kappa_{\mathrm{SR}}$ is the short-range decay constant, and $z_{\mathrm{eq}}$ is the equilibrium distance. The force-model parameters are $H_{\mathrm{A}}=10^{-20}~\mathrm{J}$, $U_0=3.71\times10^{-20}~\mathrm{J}$, $\kappa_{\mathrm{SR}}=4.255~\mathrm{nm}^{-1}$, and $z_{\mathrm{eq}}=2.35~\text{\AA}$.

For the force-distance representation, the electrostatic contribution is evaluated as a distance-dependent envelope. The quantity $C^{(1)}(z)$ is provided by the selected CG model, while the squared-bias term in Eq.~\eqref{eq:Fel} is replaced by its minimum and maximum values over one bias-modulation cycle. The electrostatic force is therefore bounded by:

\begin{subequations}\label{eq:numerical_force_el_bounds}
\begin{empheq}[left=\empheqlbrace]{align}
F_{\mathrm{el}}^{\min}(z)
&=
\frac{1}{2}C^{(1)}(z)V_{\min}^2,
\\
F_{\mathrm{el}}^{\max}(z)
&=
\frac{1}{2}C^{(1)}(z)V_{\max}^2,
\end{empheq}
\end{subequations}

\noindent with:

\begin{subequations}\label{eq:numerical_voltage_bounds}
\begin{empheq}[left=\empheqlbrace]{align}
V_{\min}^2
&=
\begin{cases}
\left(
\left|V_{\mathrm{DC}}-V_{\mathrm{cpd}}\right|
-
U_{\mathrm{mod}}
\right)^2,
&
\text{if } U_{\mathrm{mod}} < \left|V_{\mathrm{DC}}-V_{\mathrm{cpd}}\right|,
\\
0,
&
\text{if } U_{\mathrm{mod}} \geq \left|V_{\mathrm{DC}}-V_{\mathrm{cpd}}\right|,
\end{cases}
\\
V_{\max}^2
&=
\left(
\left|V_{\mathrm{DC}}-V_{\mathrm{cpd}}\right|
+
U_{\mathrm{mod}}
\right)^2.
\end{empheq}
\end{subequations}

Here, $V_{\mathrm{DC}}=0~\mathrm{V}$, $V_{\mathrm{cpd}}=+100~\mathrm{mV}$, and $U_{\mathrm{mod}}=100~\mathrm{mV}$. The bias-modulation frequency is chosen according to the heterodyne condition, $f_{\mathrm{mod}}=f_2-f_1=f_{2,0}-f_{1,0}=795~\mathrm{kHz}$.

The corresponding lower and upper bounds of the tip-surface interaction force are:

\begin{subequations}\label{eq:numerical_force_ts_bounds}
\begin{empheq}[left=\empheqlbrace]{align}
F_{\mathrm{ts}}^{\min}(z)
&=
F_{\mathrm{LR}}(z)
+
F_{\mathrm{SR}}(z)
+
F_{\mathrm{el}}^{\min}(z),
\\
F_{\mathrm{ts}}^{\max}(z)
&=
F_{\mathrm{LR}}(z)
+
F_{\mathrm{SR}}(z)
+
F_{\mathrm{el}}^{\max}(z).
\end{empheq}
\end{subequations}

The CG is computed using the Hudlet-based model introduced in Sec.~\ref{sec:Hudlet_CG_model}, including the conical, spherical-apex, and cantilever contributions. The maximum oscillatory excursion is defined as $z_{\mathrm{dyn}}=z_{1,0}$ in the monomodal case and $z_{\mathrm{dyn}}=z_{1,0}+z_{2,0}$ in the bimodal case. The interval $z_c\pm z_{\mathrm{dyn}}$ therefore defines the maximum distance range enclosed by the oscillatory motion.

\subsubsection{Force-distance landscape}

\begin{figure}[htbp]
    \centering
    \includegraphics[width=\textwidth]{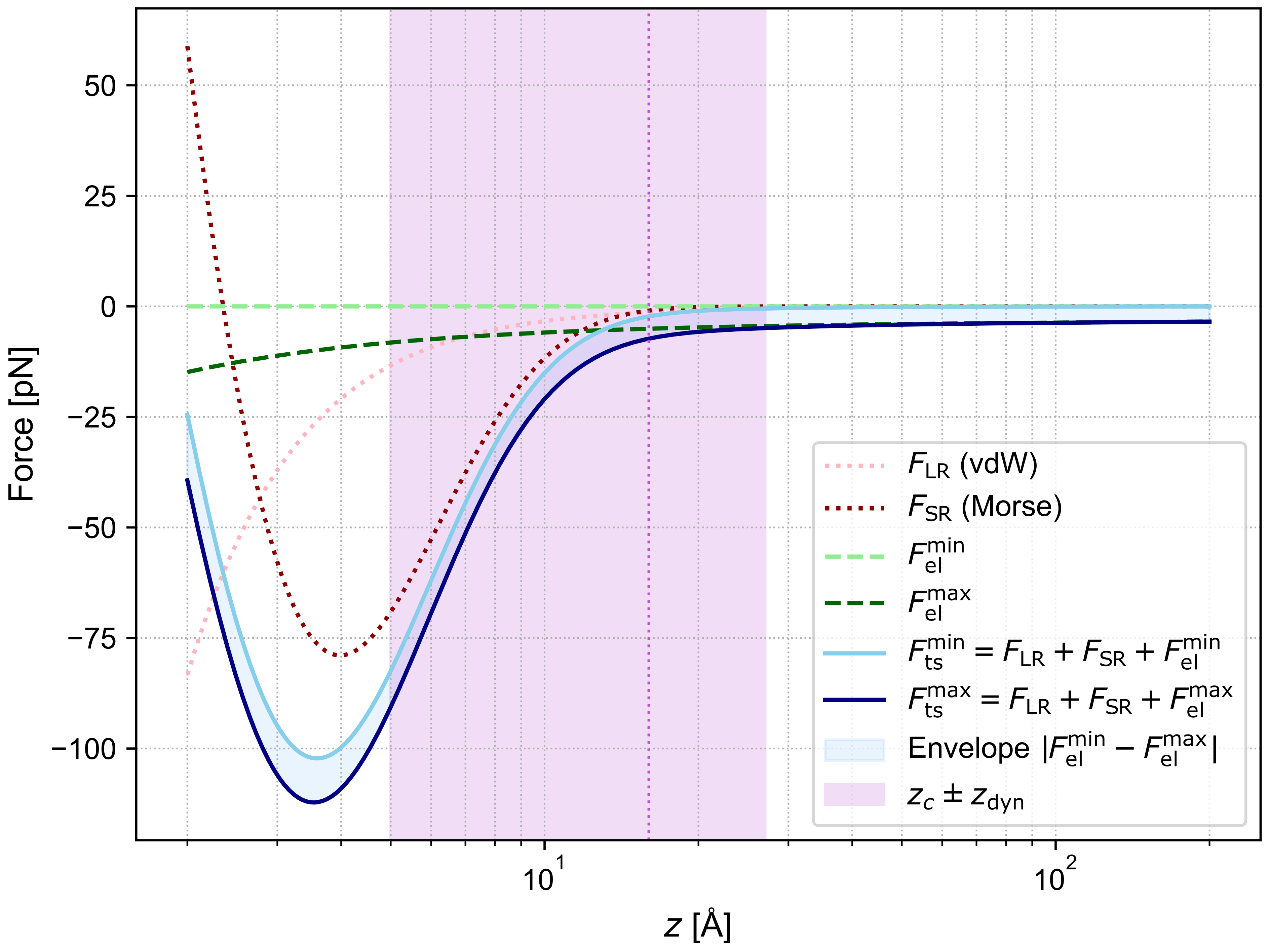}
    \caption{Numerical tip--surface interaction-force landscape displayed with a semi-logarithmic distance axis. The long-range contribution, $F_{\mathrm{LR}}$, is described by a sphere--plane van der Waals interaction, and the short-range contribution, $F_{\mathrm{SR}}$, by a Morse-like force law. The green dotted curves show the electrostatic-force bounds, $F_{\mathrm{el}}^{\min}$ and $F_{\mathrm{el}}^{\max}$, obtained from the minimum and maximum values of the squared bias over one bias-modulation cycle. The blue curves show the resulting force bounds, $F_{\mathrm{ts}}^{\min}=F_{\mathrm{LR}}+F_{\mathrm{SR}}+F_{\mathrm{el}}^{\min}$ and $F_{\mathrm{ts}}^{\max}=F_{\mathrm{LR}}+F_{\mathrm{SR}}+F_{\mathrm{el}}^{\max}$, which delimit the bias-modulation-induced force interval. The vertical purple dashed line indicates the average tip--surface distance, $z_c$, and the shaded purple region represents the sampled distance interval $z_c\pm z_{\mathrm{dyn}}$.}
    \label{fig:force_distance}
\end{figure}

The resulting force-distance curves are reported in Fig.~\ref{fig:force_distance} and exhibit typical long- and short-range variations. The semi-logarithmic distance axis emphasizes the strongly nonlinear short-distance regime while retaining the long-range part of the interaction.

The electrostatic contribution varies more smoothly with distance than the short-range force. For the selected parameters, $F_{\mathrm{el}}^{\min}=0$, whereas $F_{\mathrm{el}}^{\max}$ produces an additional attractive contribution. The separation between $F_{\mathrm{ts}}^{\min}$ and $F_{\mathrm{ts}}^{\max}$ therefore estimates the force modulation induced by the AC bias.

The vertical dashed line in Fig.~\ref{fig:force_distance} indicates the average tip-surface distance $z_c$, while the shaded region represents the interval $z_c\pm z_{\mathrm{dyn}}$ sampled during the oscillation. This interval lies entirely within the attractive branch of the tip-surface interaction force and overlaps a range where the force varies significantly and nonlinearly. The cantilever consequently samples a finite portion of the attractive force-distance curve rather than a purely local force gradient at $z_c$. The same finite-amplitude sampling governs the time dependence of $C^{(1)}(z(t))$.

\subsubsection{Time-domain response of the force}

\begin{figure}[htbp]
    \centering
    \includegraphics[width=\textwidth]{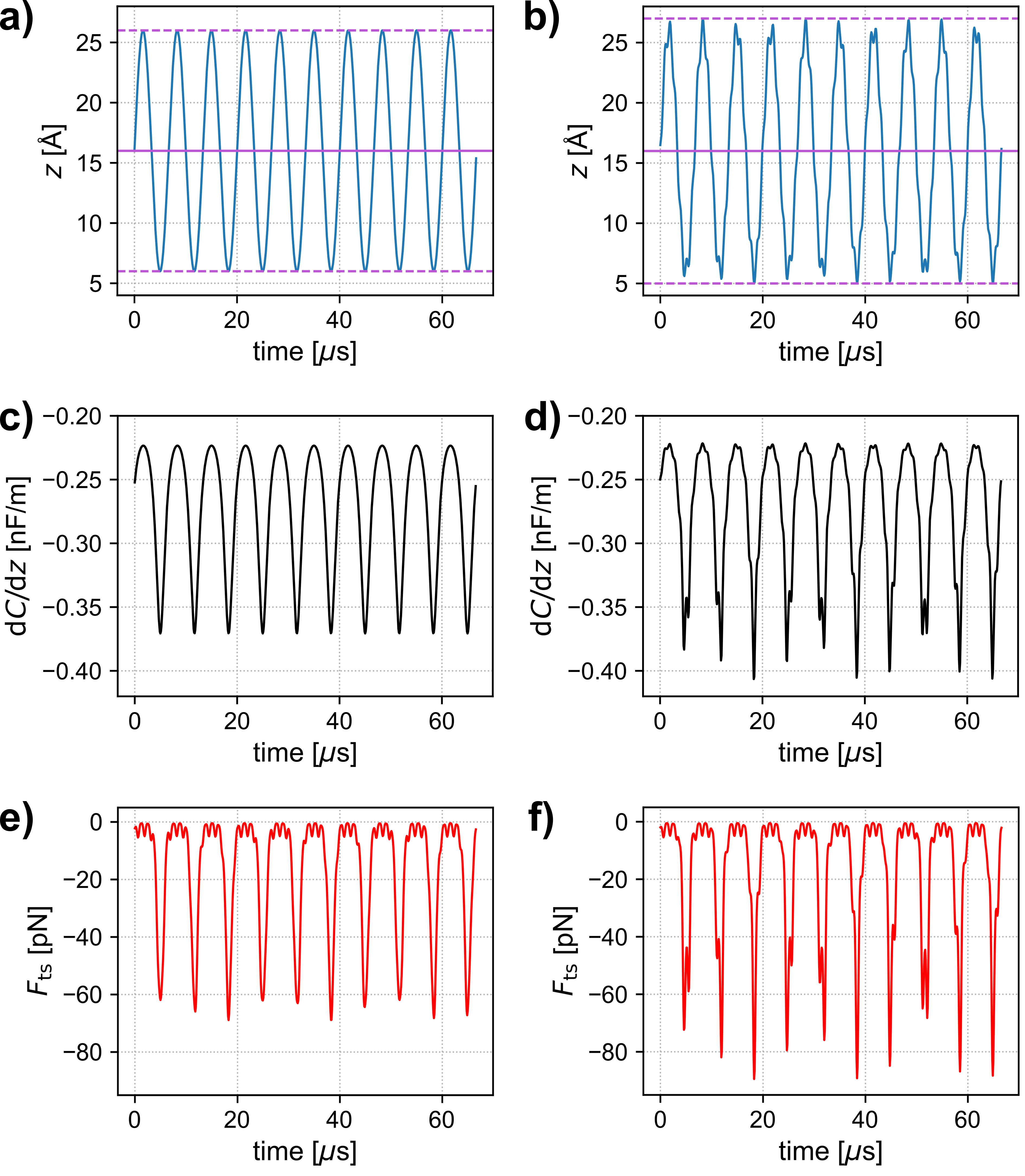}
    \caption{Time-domain numerical response of the tip--surface interaction in the monomodal and bimodal regimes. Panels (a,b) show the instantaneous tip--surface distance, $z(t)$, in the two regimes. The solid purple line indicates the average tip--surface distance, $z_c$, and the dashed purple lines the minimum and maximum sampled distances. Panels (c,d) show the corresponding CG, $C^{(1)}(z(t))$, evaluated along the tip trajectory, while panels (e,f) show the resulting force, $F_{\mathrm{ts}}(t)$, including the long-range, short-range, and electrostatic contributions. The additional second-eigenmode oscillation modifies the distance sampling and produces sharper, non-equivalent closest-approach events in both the CG and force signals.}
    \label{fig:force_time}
\end{figure}

The interaction model is then evaluated along the mechanical tip-surface distance modulation. In Fig.~\ref{fig:force_time}, the upper, middle, and lower panels display $z(t)$, $C^{(1)}(z(t))$, and the corresponding tip-surface interaction force $F_{\mathrm{ts}}(t)$, respectively. The left and right columns correspond to the monomodal and bimodal oscillations.

\begin{itemize}

\item \textbf{Monomodal oscillation.}

\end{itemize}

In the monomodal case, the tip-surface distance modulation is $T_1$-periodic. Since the CG is evaluated as a single-valued function of the instantaneous tip-surface distance, $C^{(1)}(z(t))$ is also $T_1$-periodic. Although the mechanical oscillation is sinusoidal, $C^{(1)}(z(t))$ is not, because the nonlinear dependence of $C^{(1)}$ on $z$ enhances the closest-approach part of the trajectory. This produces sharp minima in the CG signal when the tip reaches its smallest distance from the surface.

The force signal also depends on the bias modulation entering the electrostatic contribution. Its periodicity is therefore governed by the common super-frequency of the mechanical and electrical modulations. In the heterodyne configuration considered here, $f_{\mathrm{mod}}=f_2-f_1$, and the squared-bias term contains components at $f_{\mathrm{mod}}$ and $2f_{\mathrm{mod}}$. For comparison with the bimodal case, the monomodal signals in Fig.~\ref{fig:force_time} are displayed over the same temporal window, namely $10T_1$.

The phase of the dominant $f_1$ component of $C^{(1)}(z(t))$ is locked to that of $z(t)$, and the distance-dependent force components inherit the same phase-locking mechanism. This produces sharp attractive excursions of $F_{\mathrm{ts}}(t)$ at each closest-approach event.

\begin{itemize}

\item \textbf{Bimodal oscillation.}

\end{itemize}

In the bimodal case, the mechanical oscillation is governed by $f_1$ and $f_2$. For the numerical configuration considered here, $f_2=6.3f_1=63f_1/10$. The two frequencies are therefore commensurate and can be written as $f_1=10f_s$ and $f_2=63f_s$, with $f_s=f_1/10$. The corresponding super-period is:

\begin{equation}\label{eq:numerical_superperiod}
T_s=\frac{1}{f_s}=10T_1=63T_2.
\end{equation}

The bimodal oscillation $z(t)$ and the CG signal $C^{(1)}(z(t))$ are therefore $T_s$-periodic. The time window in Fig.~\ref{fig:force_time} spans $10T_1$, corresponding to one complete super-period and displaying the full repetition pattern of the bimodal distance sampling.

With $f_{\mathrm{mod}}=f_2-f_1=5.3f_1=53f_s$, the electrostatic force contains components at $f_{\mathrm{mod}}$ and $2f_{\mathrm{mod}}$ in addition to those generated by the CG dynamics. Since $f_1$, $f_2$, $f_{\mathrm{mod}}$, and $2f_{\mathrm{mod}}$ are all integer multiples of $f_s$, $F_{\mathrm{ts}}(t)$ is also periodic over $T_s=10T_1$. Its temporal structure differs from that of $C^{(1)}(z(t))$ because it results from the product of the distance-dependent CG and the modulated squared-bias term.

The $f_1$ and $f_2$ components of $C^{(1)}(z(t))$ are phase-locked to the corresponding mechanical components and therefore inherit the phases $\Phi_1$ and $\Phi_2$. The same applies to the force components generated by the distance dependence of $F_{\mathrm{ts}}(t)$. Compared with the monomodal case, the superposition of the two eigenmode oscillations makes the closest-approach events non-equivalent over the super-period. This produces a more complex temporal modulation of both $C^{(1)}(z(t))$ and $F_{\mathrm{ts}}(t)$, with sharper and more irregular force excursions.

\subsection{Capacitance gradient}
\label{sec:capacitance_gradient_numerics}

The focus is now shifted to the CG, the central quantity in the analytical developments of Sec.~\ref{sec:analytical_approach}. Starting from the Hudlet-based model, the spatial dependence of $C^{(1)}(z)$ over the distance interval explored by the tip is examined, and the respective contributions of the apex, cone, and cantilever are identified. The model is then evaluated along the mechanical trajectories to determine how its nonlinear distance dependence shapes the time-domain signal and spectral content. These numerical signals are used to assess the accuracy and convergence of the Taylor description toward the Fourier description. They are also used to determine the OTRs of the effective coefficients $K_0$, $K_1$, and $K_2$ throughout the $(z_c,z_{1,0})$ parameter space.

\subsubsection{Numerical capacitance-gradient model}

The CG is evaluated using the Hudlet-based expression introduced in Eq.~\eqref{eq:CG_Hudlet}. The geometry shown schematically in Fig.~\ref{fig:tip_cone_cantilever_geometry} consists of a rectangular cantilever. Its length is $L=200~\mu\mathrm{m}$, its width is $W=30~\mu\mathrm{m}$, and it is tilted by $\alpha_0=10^\circ$. The cone height is $H_{\mathrm{cone}}=10~\mu\mathrm{m}$, and the lever height at the tip position is taken as $H_{\mathrm{cant}}\approx H_{\mathrm{cone}}$. The cone half-aperture angle is $\theta_0=10^\circ$, and the tip-apex radius is $R=2~\mathrm{nm}$.

\begin{figure}[htbp]
    \centering
    \includegraphics[width=\textwidth]{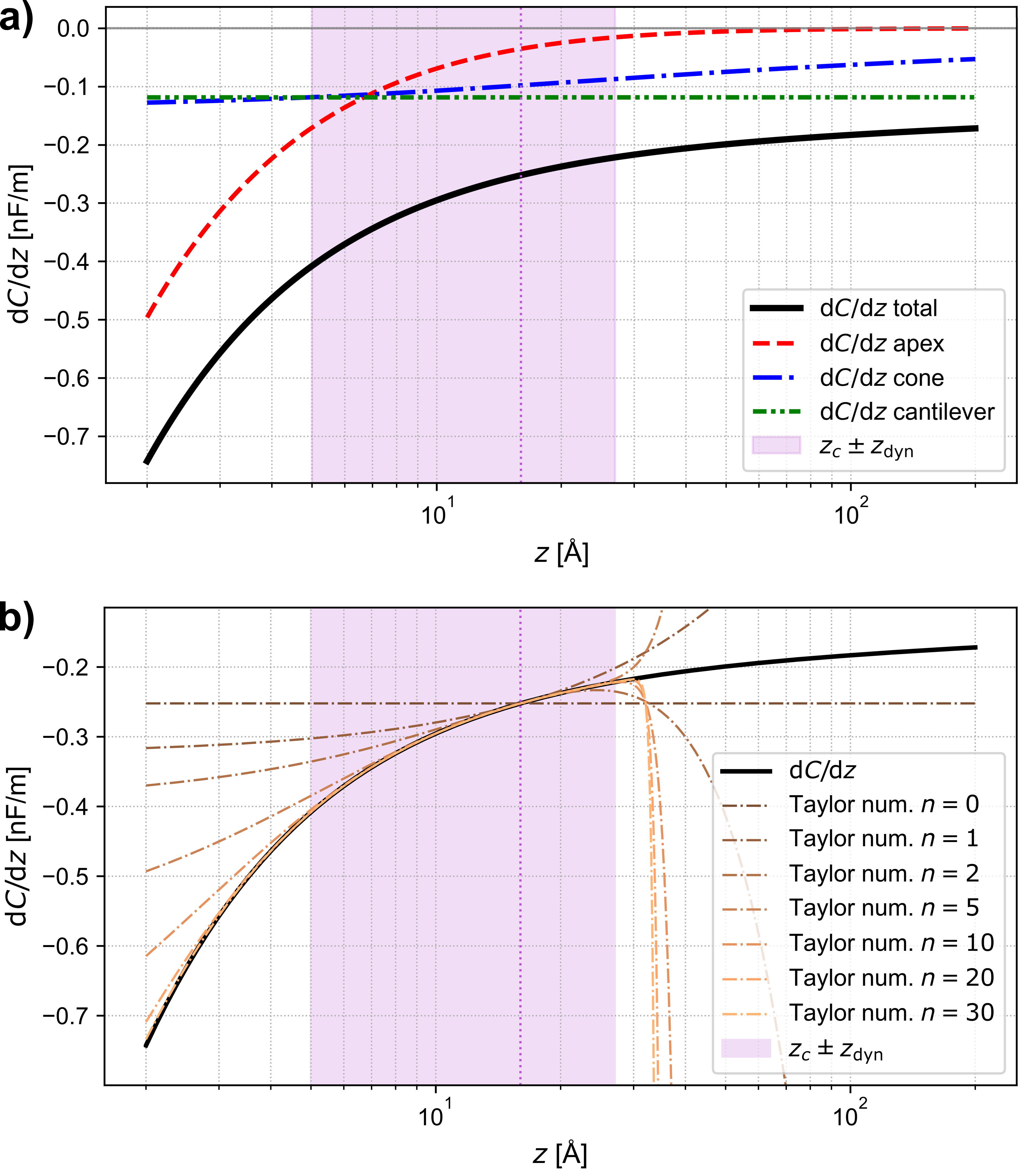}
    \caption{Numerical CG model and Taylor reconstruction in the distance domain. (a) Total CG computed from the Hudlet-based model (see Eq.~\eqref{eq:CG_Hudlet}) and its apex, cone, and cantilever contributions. The shaded purple region indicates the sampled interval $z_c\pm z_{\mathrm{dyn}}$. (b) Comparison between the exact CG, shown as a black solid curve, and Taylor reconstructions of increasing truncation order $n$ about $z_c$, shown as dash-dotted curves. Increasing $n$ improves the reconstruction over the dynamically sampled interval.}
    \label{fig:capacitance_gradient}
\end{figure}

The corresponding curves are shown in Fig.~\ref{fig:capacitance_gradient}(a). The CG is negative over the entire distance range considered. At the shortest tip-surface distances, it is dominated by the apex contribution, which exhibits the strongest distance dependence. As the distance increases, the relative contributions of the cone and cantilever become more important, leading to an intermediate regime in which all three geometrical contributions are comparable. At larger distances, the apex contribution becomes negligible and the CG is mainly governed by the cone and cantilever, with the latter eventually dominating because of its longer-range character. The distance interval sampled during the oscillation lies within the intermediate regime, where the total CG cannot be reduced to a single geometrical contribution. Consequently, the spectral components of $C^{(1)}(z(t))$ depend on the finite portion of the CG curve explored by the tip rather than on its local value at $z_c$ alone.

\subsubsection{Taylor reconstruction of the spatial capacitance gradient}

Before the time-dependent signal is considered, the Taylor reconstruction of the static function $C^{(1)}(z)$ around the average tip-surface distance $z_c$ is tested, following Eq.~\eqref{eq:CG_Taylor}. The derivatives of $C^{(1)}(z)$ at $z_c$ are computed numerically and used to construct truncated Taylor expansions of increasing order. The reconstructions are evaluated on the same logarithmic $z$-grid as the exact model, spanning $z_{\mathrm{grid}}^{\min}=2~\text{\AA}$ to $z_{\mathrm{grid}}^{\max}=20~\mathrm{nm}$ with $N_z=9800$ points.

The results are shown in Fig.~\ref{fig:capacitance_gradient}(b). Low-order expansions reproduce the local slope of $C^{(1)}(z)$ around $z_c$ but rapidly deviate from the exact curve away from the expansion point. Increasing the truncation order improves the agreement over the interval explored by the oscillating tip. The highest orders shown reproduce the exact CG throughout this dynamically sampled region, although deviations may remain outside it. The relevant criterion is therefore the reconstruction accuracy over the finite distance interval explored during the motion rather than over the entire plotted range.

\subsubsection{Spectral content in the monomodal regime}

\begin{figure}[htbp]
    \centering
    \includegraphics[width=\textwidth]{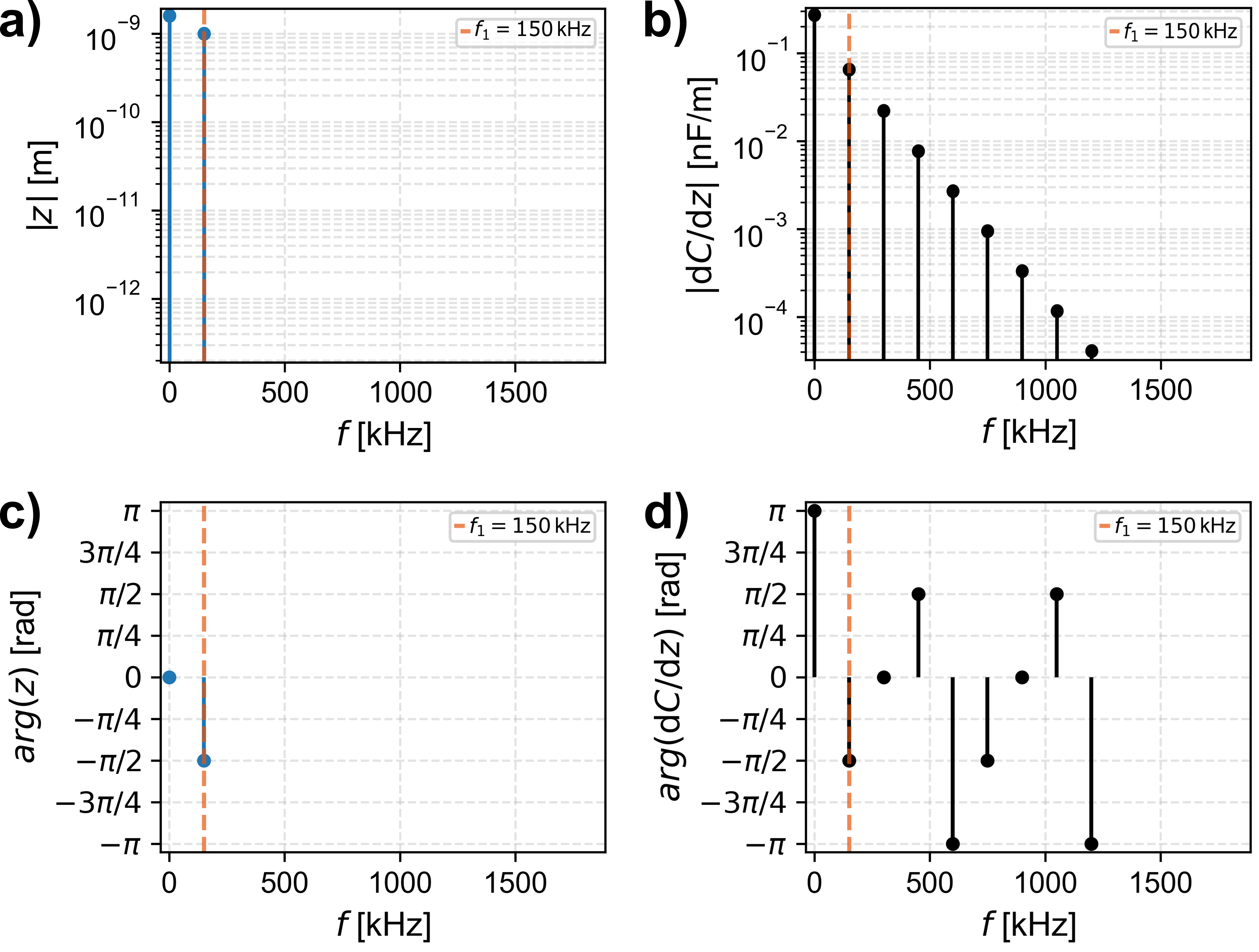}
    \caption{Fourier analysis of the monomodal CG dynamics. Panels (a,c) show the amplitude and phase spectra of the tip--surface distance signal $z(t)$, while panels (b,d) show those of $C^{(1)}(t,z_c)$. The displayed frequency range is restricted to $[0;2]~\mathrm{MHz}$ for readability, although the one-sided Nyquist interval is $[0;5]~\mathrm{MHz}$. The amplitude spectra are displayed on a semi-logarithmic $y$-axis. The mechanical motion contains only the static and $f_1$ components, the latter indicated by the orange dotted line, whereas the nonlinear dependence of $C^{(1)}(z)$ on $z(t)$ generates higher harmonics of $f_1$.}
    \label{fig:cg_fft_monomodal}
\end{figure}

The spatial nonlinearity of $C^{(1)}(z)$ directly affects the time-dependent signal $C^{(1)}(z(t))$. To quantify this effect, the prescribed mechanical motion is used to generate $z(t)$, and the exact CG signal is obtained by evaluating the Hudlet-based model along the resulting trajectory. One-sided discrete Fourier spectra are then computed for both $z(t)$ and $C^{(1)}(z(t))$. Because the signals are periodic over the analysis window, these spectra can be interpreted as numerical FSE spectra. For readability, Fig.~\ref{fig:cg_fft_monomodal} is restricted to $[0;2]~\mathrm{MHz}$. The nominal one-sided Nyquist interval is $[0;f_{\mathrm{samp}}/2]=[0;5]~\mathrm{MHz}$ for the sampling frequency reported in Table~\ref{tab:numerical_parameters}. The amplitude spectra are displayed on a semi-logarithmic $y$-axis.

The monomodal spectra are shown in Fig.~\ref{fig:cg_fft_monomodal}. The mechanical motion contains a static component and a component at $f_1$, whose amplitude matches the prescribed oscillation amplitude $z_{1,0}$ (see Fig.~\ref{fig:cg_fft_monomodal}(a)). The phase reference is the excitation force applied to the first eigenmode. Since the cantilever is driven at resonance, the $f_1$ component of $z(t)$ exhibits a phase lag of $-\pi/2$, consistently with $\Phi_1=-\pi/2$ (see Fig.~\ref{fig:cg_fft_monomodal}(c)). The zero phase of the static component reflects the positive value of the average tip-surface distance $z_c$.

Unlike the mechanical motion, the CG signal contains harmonics of $f_1$ (see Figs.~\ref{fig:cg_fft_monomodal}(b) and (d)). These harmonics are generated by the nonlinear dependence of $C^{(1)}(z)$ on the instantaneous tip-surface distance, not by additional mechanical excitations. All phases are wrapped onto the principal interval $[-\pi,\pi]$. Because the mean CG is negative, the apparent phase of its static component is $\pi$ and does not represent a dynamical phase lag. The phase of the fundamental CG component is imposed by that of the corresponding displacement component. More generally, the complete Fourier expressions derived in Sec.~II.C of the SI file give:

\begin{equation}\label{eq:monomodal_phase_relations_numerical}
\varphi_{2m}=2m\Phi_1,
\qquad
\varphi_{2m+1}=(2m+1)\Phi_1.
\end{equation}

With the present CG convention, the even-harmonic Fourier coefficients are negative, and their sign is absorbed into the phase through an additional shift of $\pi$. For $\Phi_1=-\pi/2$, the expected wrapped phases are therefore $-\pi/2$ for the fundamental and $2\Phi_1+\pi=0$ for the second harmonic. The expected wrapped phases are $3\Phi_1=\pi/2$ modulo $2\pi$ for the third harmonic and $4\Phi_1+\pi=\pi$ modulo $2\pi$ for the fourth harmonic.

\subsubsection{Spectral content in the bimodal regime}
\label{sec:spectral_content_bimodal}

\begin{figure}[htbp]
    \centering
    \includegraphics[width=\textwidth]{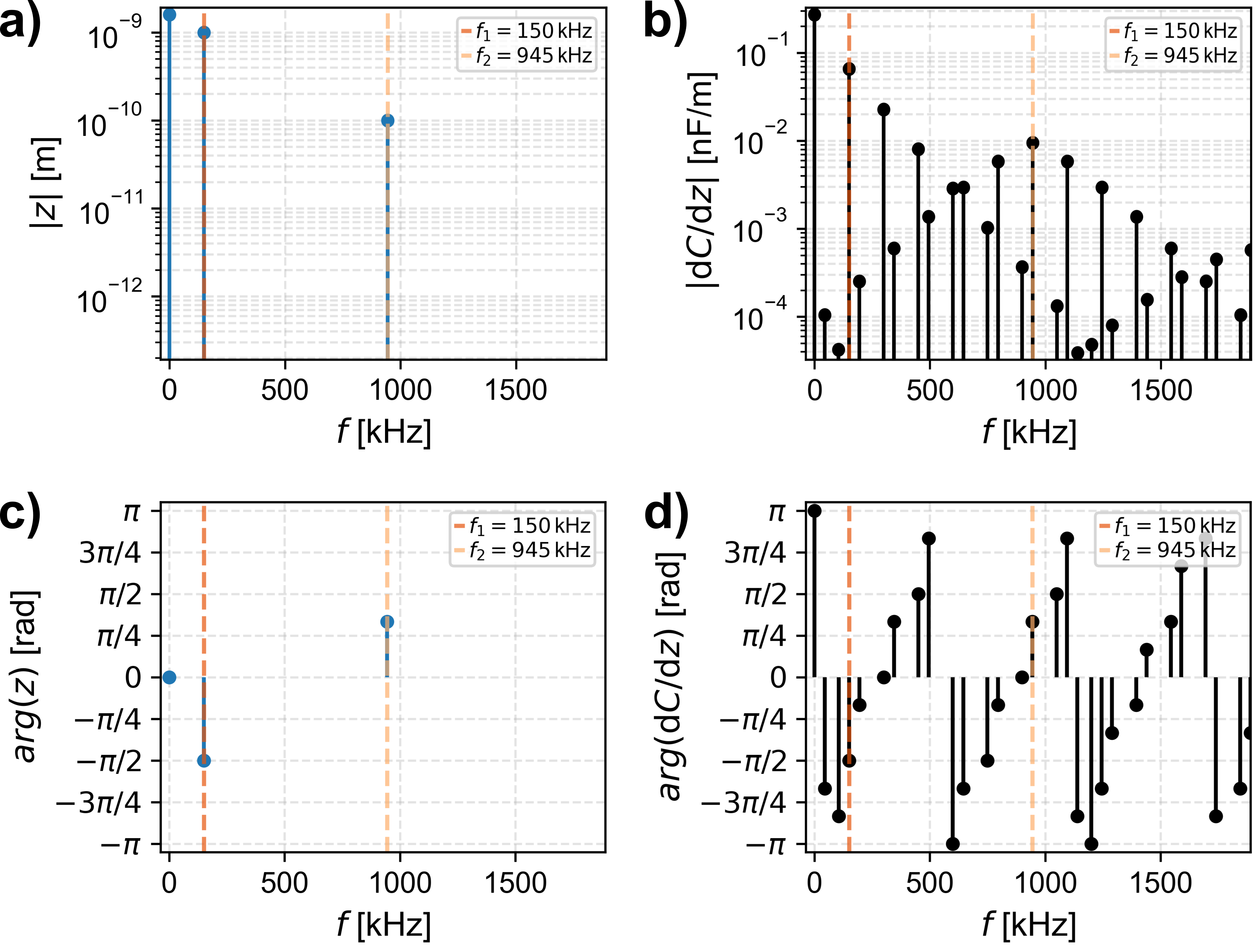}
    \caption{Fourier analysis of the bimodal CG dynamics. Panels (a,c) show the amplitude and phase spectra of the tip--surface distance signal $z(t)$, with components at $f_1$ and $f_2$, while panels (b,d) show those of $C^{(1)}(t,z_c)$. The displayed frequency range is restricted to $[0;2]~\mathrm{MHz}$ for readability, although the one-sided Nyquist interval is $[0;5]~\mathrm{MHz}$. The amplitude spectra are displayed on a semi-logarithmic $y$-axis. The nonlinear evaluation of the CG generates components at integer combinations of the two mechanical frequencies. Since $f_1=10f_s$ and $f_2=63f_s$, they lie on the harmonic comb defined by integer multiples of $f_s=15~\mathrm{kHz}$. The orange dotted lines indicate the $f_1$ and $f_2$ components, whose phases match the imposed displacement components.}
    \label{fig:cg_fft_bimodal}
\end{figure}

The same analysis is performed in the bimodal regime, where the mechanical motion contains components at $f_1$ and $f_2$. Here, $f_1=150~\mathrm{kHz}$ and $f_2=945~\mathrm{kHz}$, corresponding to $f_2/f_1=6.3$. These frequencies can be written as $f_1=10f_s$ and $f_2=63f_s$, with $f_s=15~\mathrm{kHz}$. The associated one-sided spectra are shown in Fig.~\ref{fig:cg_fft_bimodal}. As in the monomodal case, only the interval $[0;2]~\mathrm{MHz}$ is displayed, although the coefficients are calculated over the one-sided Nyquist interval $[0;5]~\mathrm{MHz}$. On the bimodal harmonic grid, this corresponds to components up to:

\begin{equation}
m_{\mathrm{max}}
=
\left\lfloor
\frac{f_{\mathrm{samp}}/2}{f_s}
\right\rfloor
=
333.
\end{equation}

The spectrum of $z(t)$ contains a static component and the two imposed mechanical components at $f_1$ and $f_2$ (see Figs.~\ref{fig:cg_fft_bimodal}(a) and (c)). Their amplitudes match $z_{1,0}$ and $z_{2,0}$, respectively. The $f_1$ component exhibits the prescribed phase lag $\Phi_1=-\pi/2$ relative to the first-eigenmode excitation force. In open-loop AM-He-KPFM, the phase of the second-eigenmode oscillation is determined by the heterodyne electrostatic drive and therefore depends on $\Phi_1$, $\Phi_{\mathrm{mod}}$, and the phase response of the second eigenmode. In the present simulations, the bimodal motion is prescribed directly and an arbitrary value $\Phi_2=\pi/3$ is used.

The spectrum of $C^{(1)}(z(t))$ is considerably richer (see Figs.~\ref{fig:cg_fft_bimodal}(b) and (d)). The nonlinear evaluation of the CG generates components at integer combinations of the two mechanical frequencies:

\begin{equation}
n_1f_1+n_2f_2,
\qquad
(n_1,n_2)\in\mathbb{Z}^2.
\end{equation}

Since $f_1$ and $f_2$ are integer multiples of $f_s$, all these components lie on the harmonic comb defined by $f_s=15~\mathrm{kHz}$. The static, $f_1$, and $f_2$ components retained in the Taylor-Fourier comparison remain clearly identifiable within this broader spectrum. Their phases are governed by $\Phi_1$ and $\Phi_2$, respectively.

Each mixed component carries a phase $n_1\Phi_1+n_2\Phi_2$ modulo $2\pi$. An additional shift of $\pi$ occurs when the sign of the corresponding Fourier coefficient is absorbed into the phase. The static CG component has no dynamical phase; its apparent phase of $\pi$ reflects only the negative sign of the DC coefficient.

This deterministic phase structure is the spectral counterpart of the TSE. Increasing the Taylor order generates higher powers of the two cosine terms associated with $f_1$ and $f_2$. The resulting cosine-product identities produce components at integer combinations of the mechanical frequencies and phases formed from the corresponding combinations of $\Phi_1$ and $\Phi_2$. The phase distribution in Fig.~\ref{fig:cg_fft_bimodal}(d) therefore reflects the Taylor-Fourier correspondence derived in Sec.~\ref{sec:analytical_approach} and detailed in Sec.~III.B of the SI file.

\subsubsection{Time-domain reconstruction from Taylor and Fourier descriptions}

The convergence between the TSE and FSE descriptions is also examined in the time domain. The exact signal $C^{(1)}(t,z_c)$ obtained from Eq.~\eqref{eq:CG_Hudlet} is compared with two reconstructions. The first is a Taylor reconstruction about $z_c$, evaluated along $z(t)-z_c$. The second is a Fourier reconstruction obtained by harmonic synthesis from the one-sided discrete spectrum of the exact signal. The signals are sampled at $f_{\mathrm{samp}}=10~\mathrm{MHz}$ over $T_w=0.1~\mathrm{s}$. This corresponds to $N_{\mathrm{samp}}=10^6$ samples and a spectral resolution $\delta f=1/T_w=10~\mathrm{Hz}$. In the bimodal case, the reconstruction uses the harmonic grid associated with $f_s=15~\mathrm{kHz}$ up to the Nyquist frequency, corresponding to $m_{\mathrm{max}}=333$ harmonics.

\begin{figure}[htbp]
    \centering
    \includegraphics[width=\textwidth]{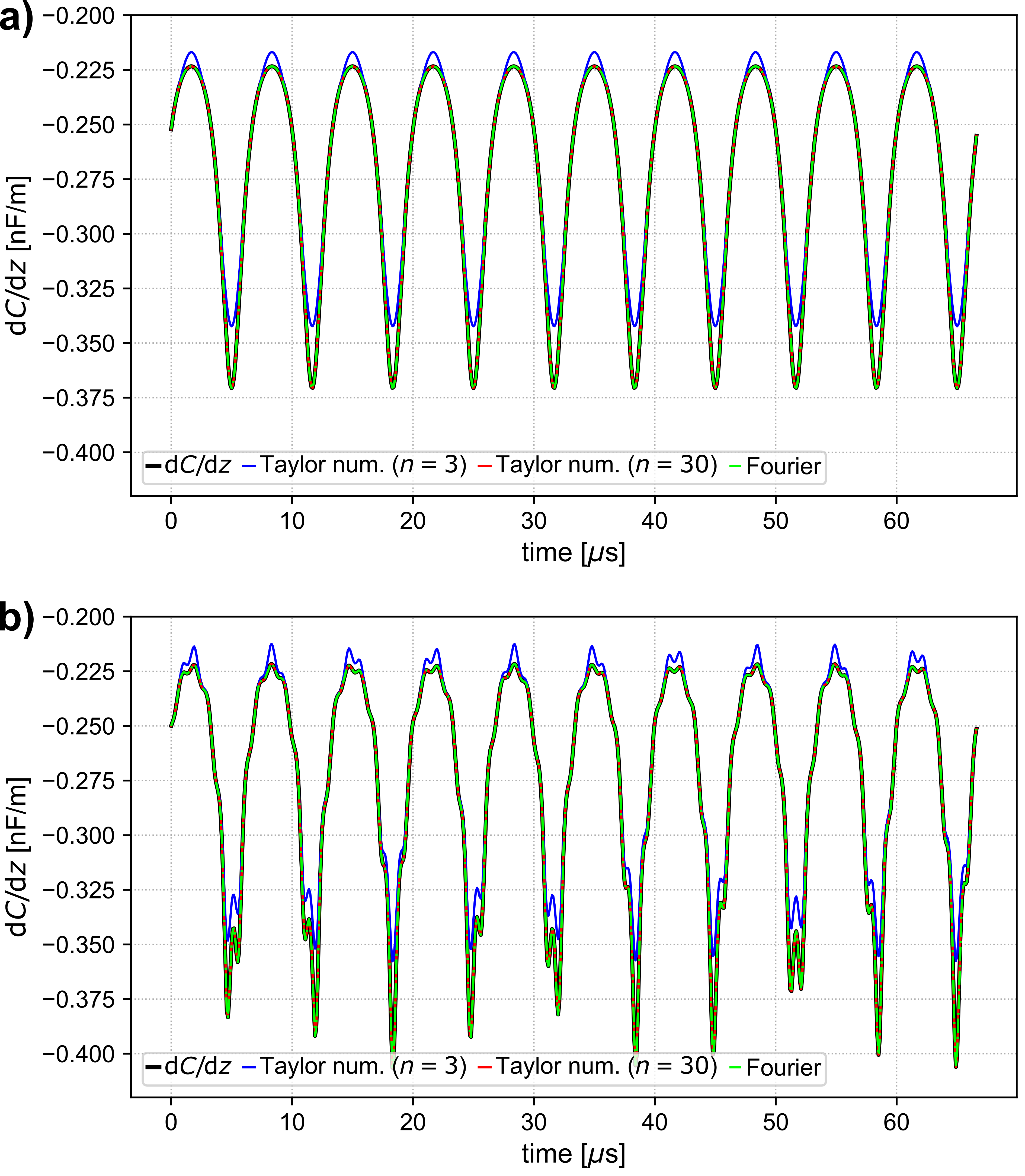}
    \caption{Time-domain reconstruction of the CG dynamics from Taylor and Fourier descriptions. (a) Monomodal case. (b) Bimodal case over one super-period. The exact signal $C^{(1)}(t,z_c)$, computed from the Hudlet-based model in Eq.~\eqref{eq:CG_Hudlet}, is compared with the Fourier reconstruction obtained by harmonic synthesis. It is also compared with Taylor reconstructions from Eq.~\eqref{eq:CG_Taylor} at truncation orders $n=3$ and $n=30$. The exact signal is shown in black, the Fourier reconstruction as a green dotted curve, and the Taylor reconstructions for $n=3$ and $n=30$ in blue and red, respectively. The low-order reconstruction captures the global modulation but not the sharp closest-approach events, whereas the exact, Fourier, and high-order Taylor curves overlap with excellent agreement.}
    \label{fig:cg_time_reconstruction}
\end{figure}

The comparison is shown in Fig.~\ref{fig:cg_time_reconstruction}, and the corresponding errors are reported in Table~\ref{tab:time_reconstruction_errors}. In the monomodal case (see Fig.~\ref{fig:cg_time_reconstruction}(a)), the third-order Taylor expansion captures the global modulation but does not accurately reproduce the sharp extrema associated with closest approach. Over one first-eigenmode period, the normalized root-mean-square and maximum absolute errors are $6.80\%$ and $19.2\%$, respectively. These extrema correspond to the strongest variations of $C^{(1)}(t,z_c)$ and require higher-order derivatives. At $n=30$, the errors decrease to $1.30\times10^{-5}\%$ and $5.90\times10^{-5}\%$, respectively.

The same behavior is observed in the bimodal case (see Fig.~\ref{fig:cg_time_reconstruction}(b)), with more pronounced low-order deviations because the closest-approach events are not equivalent over the super-period. At $n=3$, the normalized root-mean-square and maximum absolute errors over one complete super-period are $6.28\%$ and $26.4\%$, respectively. At $n=30$, they decrease to $1.02\times10^{-4}\%$ and $9.75\times10^{-4}\%$. The exact, Fourier-reconstructed, and high-order Taylor-reconstructed signals then overlap with excellent accuracy, confirming the equivalence of the two descriptions at sufficiently high Taylor order.

\subsubsection{Convergence of the Taylor coefficients toward the Fourier coefficients}

The convergence of the Taylor-based coefficients toward the Fourier coefficients is tested using the exact CG signal from Eq.~\eqref{eq:CG_Hudlet}, evaluated along the prescribed bimodal trajectory. The reference Fourier coefficients associated with the static, $f_1$, and $f_2$ components are obtained by projection over the analysis window. The static coefficient is the time average of the exact signal, while the first- and second-eigenmode coefficients are obtained by projection onto:

\begin{equation}
\cos(2\pi f_1t+\Phi_1)
\qquad\text{and}\qquad
\cos(2\pi f_2t+\Phi_2),
\end{equation}

\noindent respectively. These coefficients are denoted $X_0$, $X_{p_1}$, and $X_{p_2}$.

For each Taylor truncation order $n$, the CG is approximated by the TSE of $C^{(1)}(z)$ about $z_c$. This approximation is evaluated along $z(t)-z_c$, yielding the truncated signal $C_n^{(1)}(t,z_c)$. The numerical Taylor-based estimates are provided by applying the same projection procedure:

\begin{equation}
K_0^{\mathrm{num}}(n),
\qquad
K_1^{\mathrm{num}}(n)z_{1,0},
\qquad
K_2^{\mathrm{num}}(n)z_{2,0}.
\end{equation}

These estimates are compared with $X_0$, $X_{p_1}$, and $X_{p_2}$, respectively.

A second estimate is obtained directly from the analytical Taylor-Fourier expressions derived in Sec.~\ref{sec:analytical_approach}. It uses the CG derivatives at $z_c$, the oscillation amplitudes, and the eigenfrequency relationship between the two modes (see Eqs.~\eqref{eq:bimodal_Taylor_prominent_components}-\eqref{eq:bimodal_Lambda}). These coefficients are denoted:

\begin{equation}
K_0^{\mathrm{ana}}(n),
\qquad
K_1^{\mathrm{ana}}(n)z_{1,0},
\qquad
K_2^{\mathrm{ana}}(n)z_{2,0}.
\end{equation}

The numerical projections test the convergence of the truncated Taylor signal, whereas the analytical estimates test the closed-form Taylor-Fourier coefficients.

The analysis is presented for the bimodal case because it contains all components of the general formulation. The monomodal representation differs only through the absence of $X_{p_2}$, $K_2z_{2,0}$, and the associated errors $\varepsilon_2^{\mathrm{num}}$ and $\varepsilon_2^{\mathrm{ana}}$.

\begin{figure}[htbp]
    \centering
    \includegraphics[width=\textwidth]{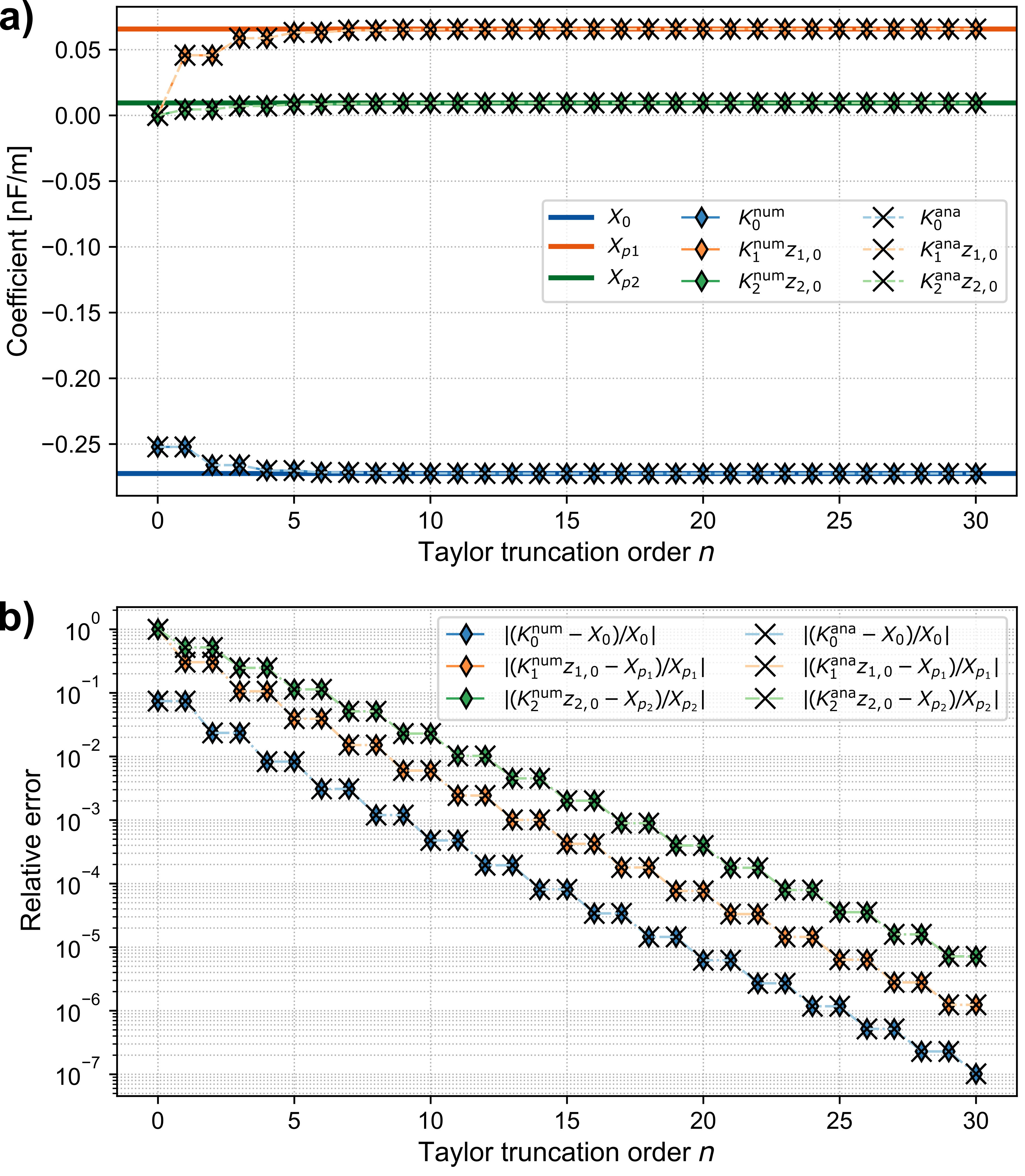}
    \caption{Convergence of the Taylor-based CG coefficients toward the Fourier coefficients in the bimodal case. (a) Evolution of $K_0$, $K_1z_{1,0}$, and $K_2z_{2,0}$ with Taylor truncation order $n$, compared with the Fourier coefficients $X_0$, $X_{p_1}$, and $X_{p_2}$. The Taylor-based coefficients are obtained from Eqs.~\eqref{eq:bimodal_Taylor_Coefs}--\eqref{eq:bimodal_TSE_FSE}. The superscripts ``num'' and ``ana'' denote values obtained by numerical projection of the truncated Taylor signal and from the analytical Taylor--Fourier expressions, respectively. (b) Corresponding relative errors, defined in Eq.~\eqref{eq:relative_errors}. Convergence toward the Fourier description is confirmed by the staircase-like decrease of the errors. The monomodal case is not shown because it differs only by the absence of $X_{p_2}$, $K_2z_{2,0}$, and the associated errors.}
    \label{fig:cg_coeff_convergence}
\end{figure}

The results are shown in Fig.~\ref{fig:cg_coeff_convergence}. As $n$ increases, the Taylor-based quantities converge toward their Fourier counterparts (see Fig.~\ref{fig:cg_coeff_convergence}(a)). Specifically, $K_0$, $K_1z_{1,0}$, and $K_2z_{2,0}$ converge toward $X_0$, $X_{p_1}$, and $X_{p_2}$, respectively. The numerical and analytical estimates overlap over the complete range of truncation orders, confirming that the closed-form expressions reproduce the spectral components extracted from the numerical signals.

The staircase-like evolution of the coefficients follows from the parity correspondence summarized in Table~\ref{tab:OTR_Taylor_order_correspondence}. The static coefficient $K_0$ is modified only by even-order Taylor terms, whereas $K_1z_{1,0}$ and $K_2z_{2,0}$ are modified only by odd-order terms. Each coefficient therefore remains unchanged when a Taylor term of the opposite parity is introduced.

For $\chi\in\{\mathrm{num},\mathrm{ana}\}$, the relative errors shown in Fig.~\ref{fig:cg_coeff_convergence}(b) are defined as:

\begin{subequations}
\label{eq:relative_errors}
\begin{empheq}[left=\empheqlbrace]{align}
\varepsilon_0^{\chi}(n)
&=
\left|
\frac{K_0^{\chi}(n)-X_0}
{X_0}
\right|,
\\
\varepsilon_1^{\chi}(n)
&=
\left|
\frac{K_1^{\chi}(n)z_{1,0}-X_{p_1}}
{X_{p_1}}
\right|,
\\
\varepsilon_2^{\chi}(n)
&=
\left|
\frac{K_2^{\chi}(n)z_{2,0}-X_{p_2}}
{X_{p_2}}
\right|.
\end{empheq}
\end{subequations}

These Fourier-referenced errors are not used for the analytical OTR attribution because the Fourier coefficients are obtained from numerical projections of the exact signal. They nevertheless provide an independent validation of the Taylor-Fourier correspondence and quantify the spectral error associated with each truncation order.

The errors decrease exponentially with $n$, consistently with the time-domain reconstruction. At $n=3$, the relative errors of the $f_1$ and $f_2$ coefficients remain of the order of $10^{-1}$, explaining the visible deviations from the exact signal. At $n=30$, they decrease to approximately $10^{-6}$. This validates the Taylor-Fourier correspondence for the selected numerical configuration.

\subsubsection{Numerical attribution of the order-truncation regimes}
\label{sec:numerical_OTR_maps}

The analytical term-significance criterion introduced in Sec.~\ref{sec:bimodal_OTR} is applied numerically to map the OTRs of $K_0$, $K_1$, and $K_2$. The objective is to identify the highest grouped Taylor order that contributes significantly to each coefficient within a representative experimental parameter space.

For each $K_i$, the relative contribution of the grouped term $T_{i,m}$ is evaluated with respect to the leading term $T_{i,0}$ through:

\begin{equation}\label{eq:numerical_OTR_significance_ratio}
r_{i,m}
=
\left|
\frac{T_{i,m}}{T_{i,0}}
\right|.
\end{equation}

A grouped order $m$ is significant when $r_{i,m}\geq\tau$, and the assigned OTR order is the largest value satisfying this condition (see Eq.~\eqref{eq:monomodal_OTR_order}). The grouped terms and attribution conditions are given for the monomodal and bimodal cases in Secs.~II.E and III.F of the SI file, respectively. The bimodal grouped terms from Sec.~III.F of the SI file are used in the present maps.

The significance threshold is set to $\tau=10^{-2}$. A correction is therefore retained when its magnitude reaches at least $1\%$ of the leading term of the same $K_i$ series. The relative significance level adopted for the numerical attribution is defined by this working value, but it should not be interpreted as a universal physical boundary between OTRs.

The OTRs are evaluated in the $(z_c,z_{1,0})$ plane for a fixed second-eigenmode amplitude $z_{2,0}=0.1~\mathrm{nm}$. Both axes are logarithmic to cover several decades of average tip-surface distance and first-eigenmode amplitude. The domain is restricted to physically admissible trajectories satisfying:

\begin{equation}\label{eq:numerical_OTR_physical_condition}
z_{\min}
=
z_c-z_{1,0}-z_{2,0}
>
0.
\end{equation}

The excluded region is shown in gray, its boundary is indicated by the black dashed line, and the red dashed curves delimit the successive OTR domains. The classification distinguishes ZOTR, FOTR, HOTR-2, HOTR-3, HOTR-4, HOTR-5, and HOTR-\textgreater{}5, the latter grouping all points for which $\ell_i^\star>5$.

\begin{figure}[htbp]
    \centering
    \includegraphics[width=\textwidth]{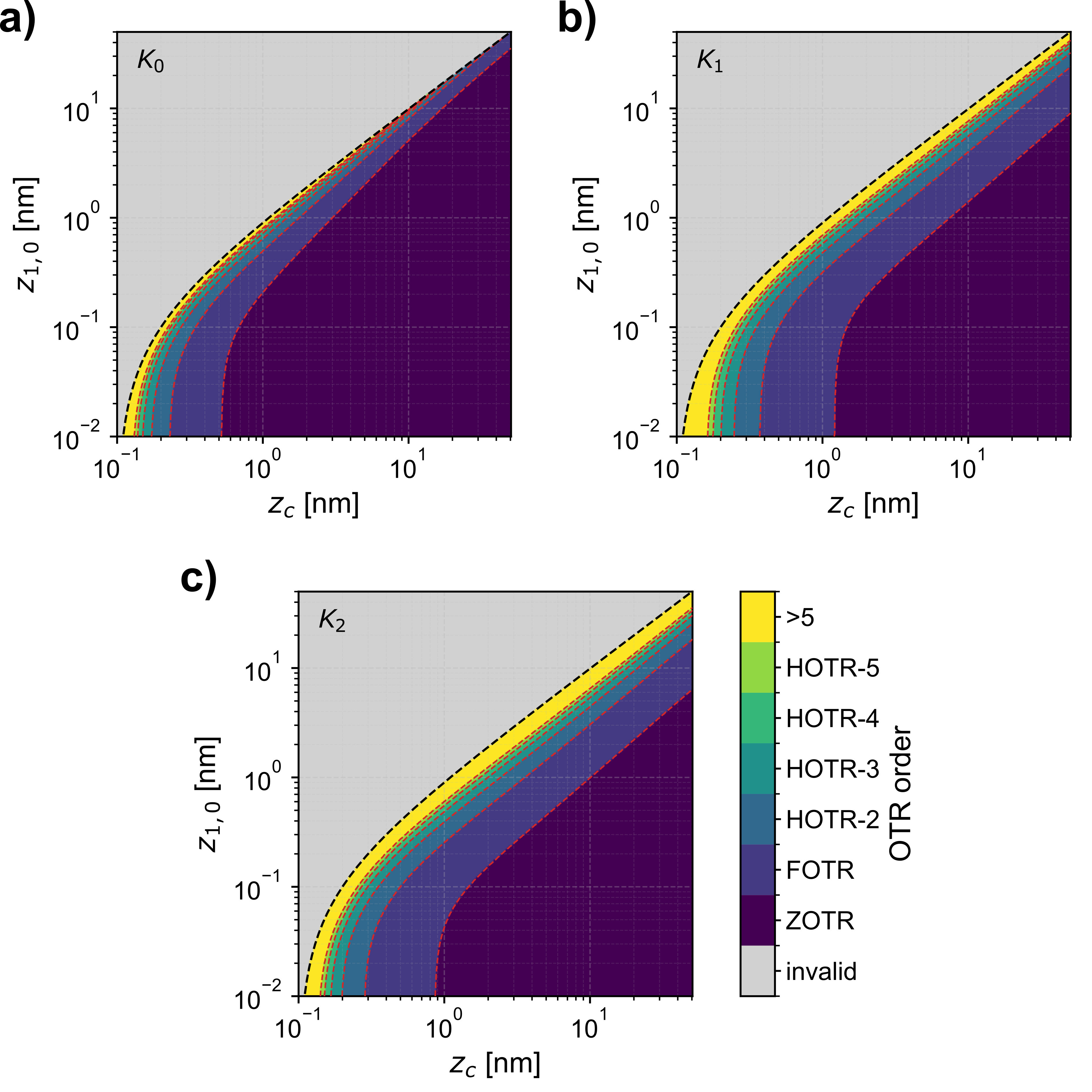}
    \caption{Numerical attribution of the order-truncation regimes in the $(z_c,z_{1,0})$ plane for $z_{2,0}=0.1~\mathrm{nm}$ and $\tau=10^{-2}$. Panels (a), (b), and (c) correspond to $K_0$, $K_1$, and $K_2$, respectively. The assigned OTR order is the largest grouped order $m$ satisfying $|T_{i,m}/T_{i,0}|\geq\tau$ (see Eq.~\eqref{eq:numerical_OTR_significance_ratio}); the grouped terms are given in Sec.~III.F of the SI file. The purple domains correspond to ZOTR and FOTR, and the blue-to-yellow domains to HOTR-2, HOTR-3, HOTR-4, HOTR-5, and HOTR-\textgreater{}5. Red dashed curves delimit successive OTR domains. The gray region corresponds to mechanically inadmissible trajectories satisfying $z_c-z_{1,0}-z_{2,0}\leq0$, and the black dashed line marks the boundary $z_c-z_{1,0}-z_{2,0}=0$. A component-dependent OTR attribution is shown by the comparison, and $K_2$ is shown to generally enter higher-order regimes before $K_1$.}
    \label{fig:otr_maps}
\end{figure}

The resulting maps are shown in Fig.~\ref{fig:otr_maps}. For all three coefficients, ZOTR occupies the region of small oscillation amplitudes relative to the average tip-surface distance. Within this domain, the leading terms $C^{(1)}(z_c)$ for $K_0$ and $C^{(2)}(z_c)$ for $K_1$ and $K_2$ are sufficient at the prescribed threshold. Increasing $z_{1,0}$ at fixed $z_c$, or equivalently decreasing the minimum sampled distance, drives the system successively toward FOTR and higher-order regimes. This occurs because higher-order spatial derivatives of the CG become increasingly important as the probe samples a broader and more nonlinear distance range.

The OTR boundaries differ between the three coefficients because the amplitudes enter their grouped corrections with different combinatorial weights. In particular, $K_2$ enters higher-order regimes before $K_1$ over a substantial part of the parameter space. This is consistent with the analytical limit $z_{1,0}\gg z_{2,0}$, for which the first correction to $K_2$ is twice that to $K_1$ (see Eq.~\eqref{eq:bimodal_FOTR_correction_ratio}). The second-eigenmode CG component may therefore require a higher truncation order even when $z_{2,0}$ remains small.

A direct numerical interpretation of the analytical OTR criterion is provided by these maps. A geometry-independent ``small-amplitude'' condition is thus replaced by component-dependent boundaries. These boundaries are determined jointly by $z_c$, $z_{1,0}$, $z_{2,0}$, the selected capacitance model, the complete tip-surface geometry, and the prescribed significance threshold.

\clearpage

\section{Conclusion}\label{sec:conclusion}

In this work, a rigorous analytical framework has been established for describing capacitance-gradient dynamics in heterodyne Kelvin probe force microscopy. Its foundation is a non-truncated Taylor-series representation of the time-dependent tip-surface capacitance gradient (see Eq.~\eqref{eq:CG_Taylor}). This representation is formulated independently of the number of mechanical components (Sec.~\ref{sec:general_Taylor_description}). It is also shown to converge for a realistic Hudlet-based tip-surface capacitance model along physically admissible monomodal and bimodal trajectories (Sec.~\ref{sec:Taylor_series_convergence}). The framework assumes prescribed sinusoidal mechanical trajectories and a quasi-static description of the tip-surface electrostatic interaction. Whenever the motion is periodic, the Taylor-series representation can be directly compared with a Fourier-series description. It thereby resolves a central limitation of the standard first-order treatment used in heterodyne KPFM, namely its implicit restriction to a vaguely defined low-amplitude regime.

In the monomodal case, explicit expressions for the dominant Fourier coefficients of the capacitance gradient are obtained from the formalism. An analytical term-significance criterion is also introduced for assigning order-truncation regimes. It replaces the qualitative notion of ``small-oscillation-amplitude regime''. With this criterion, it is established when the leading-order approximation is accurate within a prescribed tolerance and when higher-order corrections must be retained. It is also shown that increasing the oscillation amplitude or decreasing the tip-surface distance enhances the sensitivity of the dynamics to short-range electrostatic contributions contained in higher-order capacitance-gradient derivatives. The convergence of the Taylor-based coefficients toward their Fourier counterparts is confirmed by numerical simulations, and component-dependent order-truncation-regime maps are provided in the average tip-surface distance and first-eigenmode amplitude parameter space. These maps show that the system is driven toward higher-order regimes by increasing the first-eigenmode oscillation amplitude or decreasing the average tip-surface distance.

The same framework has been extended to the bimodal case, in which the capacitance-gradient dynamics is governed by the joint oscillation of two cantilever eigenmodes. The resulting spectral content consists of components at integer combinations of the two eigenmode frequencies. For commensurate frequencies, these components lie on the harmonic comb associated with the bimodal super-period. Incommensurate frequencies instead produce a quasi-periodic discrete spectrum without a finite super-period (see Sec.~III.A of the supplementary information file). The relevant static, first-eigenmode, and second-eigenmode coefficients can be expressed analytically in terms of higher-order capacitance-gradient derivatives. This treatment is essential for open-loop amplitude-modulation heterodyne KPFM, where the second eigenmode reaches a finite amplitude and contributes to the tip-surface distance modulation. The analytical coefficients governing the electrostatic-force components in this regime are therefore provided by the present work.

The present manuscript and its companion study consequently play distinct but complementary roles. In the current work, the analytical description of the capacitance-gradient dynamics is established and numerically validated, and the corresponding term-significance-based order-truncation regimes are introduced. In the companion manuscript, this foundation is used to derive the direct and inverse heterodyne force components acting on the two eigenmodes. The inverse heterodyne effect and the associated inter-mode energy exchange are also experimentally demonstrated. It is further shown how the capacitance-gradient nonlinearity captured by the different order-truncation regimes affects the observables measured in open-loop amplitude-modulation heterodyne KPFM. 

\clearpage

\section*{Acknowledgments}\label{sec:acknowledgments}

This work was supported by the Centre National de la Recherche Scientifique (CNRS) and Aix-Marseille Université. The authors thank the ANR funding agency for financial support of the PESOS project (ANR-23-CE09-0038, H.V., S.C., C.L., L.N. and B.G.) and the SuperZIC project (ANR-22-CE09-0020, S.C., C.L. and L.N.).

\section*{Author Contributions}\label{sec:author-contributions}

H.V.: conceptualization; methodology; software; data curation; formal analysis; investigation; validation; writing---original draft preparation; writing---review and editing. S.C.: investigation; validation; writing---review and editing. C.L.: investigation; validation; writing---review and editing. L.N.: conceptualization; methodology; data curation; formal analysis; investigation; validation; writing---original draft preparation; writing---review and editing; funding acquisition; supervision. B.G.: conceptualization; methodology; formal analysis; investigation; validation; writing---original draft preparation; writing---review and editing; funding acquisition; supervision; project administration.

\section*{Data Availability Statement}\label{sec:data-availability-statement}

The data that support the findings of this study are available from the corresponding author upon reasonable request.

\clearpage


\makeatletter
\renewcommand{\bibsection}{%
  \section*{References}%
}
\makeatother

\bibliographystyle{apsrev4-2}
\bibliography{uhvafm_bib}

\clearpage

\section*{Figures}\label{sec:figures}

\clearpage

\section*{Tables}\label{sec:tables}

\begin{table}[htbp]
\centering
\caption{Correspondence between the grouped levels used for OTR attribution and the truncation order $n$ of the original Taylor-series expansion. The indices $q$ and $m$ apply to the monomodal and bimodal coefficients, respectively. An OTR order $\ell$ therefore does not generally correspond to $n=\ell$. ZOTR retains the contributions originating from $n=0$ for $K_0$ and $n=1$ for $K_1$ and $K_2$. FOTR additionally retains those from $n=2$ and $n=3$, respectively. HOTR-$\ell$ extends the truncation to $n=2\ell$ for $K_0$ and $n=2\ell+1$ for $K_1$ and $K_2$.}
\label{tab:OTR_Taylor_order_correspondence}
\scriptsize
\renewcommand{\arraystretch}{1.2}
\begin{tabular}{|c|c|c|c|}
\hline
\textbf{Coefficient}
&
\textbf{Monomodal grouped level}
&
\textbf{Bimodal grouped level}
&
\textbf{Corresponding Taylor order}
\\
\hline
$K_0$
&
$q$
&
$m$
&
$n=2q$ or $n=2m$
\\
\hline
$K_1$
&
$q$
&
$m$
&
$n=2q+1$ or $n=2m+1$
\\
\hline
$K_2$
&
Not applicable
&
$m$
&
$n=2m+1$
\\
\hline
\end{tabular}
\end{table}

\begin{table}[htbp]
\centering
\caption{Numerical parameters used in the simulations.}
\label{tab:numerical_parameters}
\scriptsize
\renewcommand{\arraystretch}{1.0}
\begin{adjustbox}{max width=\textwidth, max totalheight=0.8\textheight, keepaspectratio}
\begin{tabular}{|l|c|c|}
\hline
\textbf{Cantilever dynamics and stiffness:} & \textbf{Eigenmode 1} & \textbf{Eigenmode 2} \\
\hline
Resonance frequency
& $f_{1,0}=150~\mathrm{kHz}$
& $f_{2,0}=6.3\,f_{1,0}=945~\mathrm{kHz}$ \\
\hline
Super-frequency
& \multicolumn{2}{c|}{$f_s=15~\mathrm{kHz}$} \\
\hline
Super-period
& \multicolumn{2}{c|}{$T_s=1/f_s\approx 66.6~\mu\mathrm{s}$} \\
\hline
Stiffness
& $k_1=48~\mathrm{N/m}$
& $k_2=39.3\,k_1\approx 1886~\mathrm{N/m}$ \\
\hline
Oscillation amplitude
& $z_{1,0}=1~\mathrm{nm}$
& $\begin{array}{l}
z_{2,0}=0~\mathrm{nm}\ \text{(monomodal)}\\
z_{2,0}=0.1~\mathrm{nm}\ \text{(bimodal)}
\end{array}$ \\
\hline
Phase
& $\Phi_1=-\pi/2$
& $\Phi_2=\pi/3$ \\
\hline
Average tip--surface distance
& \multicolumn{2}{c|}{$z_c=1.6~\mathrm{nm}$} \\
\hline
Minimum tip--surface distance
& \multicolumn{2}{c|}{$
\begin{array}{l}
z_{\min}=z_c-z_{1,0}=0.6~\mathrm{nm}\ \text{(monomodal)}\\
z_{\min}=z_c-z_{1,0}-z_{2,0}=0.5~\mathrm{nm}\ \text{(bimodal)}
\end{array}
$} \\
\hline

\multicolumn{3}{|l|}{\textbf{Cantilever geometry:}} \\
\hline
Cantilever:
& \multicolumn{2}{c|}{} \\
\hline
Length
& \multicolumn{2}{c|}{$L=200~\mu\mathrm{m}$} \\
\hline
Width
& \multicolumn{2}{c|}{$W=30~\mu\mathrm{m}$} \\
\hline
Lever height at the tip position
& \multicolumn{2}{c|}{$H_{\mathrm{cant}}\approx H_{\mathrm{cone}}=10~\mu\mathrm{m}$} \\
\hline
Tilt angle
& \multicolumn{2}{c|}{$\alpha_0=10^\circ$} \\
\hline
Cone:
& \multicolumn{2}{c|}{} \\
\hline
Cone height
& \multicolumn{2}{c|}{$H_{\mathrm{cone}}=10~\mu\mathrm{m}$} \\
\hline
Cone half-aperture angle
& \multicolumn{2}{c|}{$\theta_0=10^\circ$} \\
\hline
Tip:
& \multicolumn{2}{c|}{} \\
\hline
Tip-apex radius
& \multicolumn{2}{c|}{$R=2~\mathrm{nm}$} \\
\hline

\multicolumn{3}{|l|}{\textbf{Tip--surface interaction force:}} \\
\hline
Long-range and short-range interactions:
& \multicolumn{2}{c|}{} \\
\hline
Hamaker constant
& \multicolumn{2}{c|}{$H_{\mathrm{A}}=10^{-20}~\mathrm{J}$} \\
\hline
Potential depth
& \multicolumn{2}{c|}{$U_0=3.71\times10^{-20}~\mathrm{J}$} \\
\hline
Short-range decay constant
& \multicolumn{2}{c|}{$\kappa_{\mathrm{SR}}=4.255~\mathrm{nm}^{-1}$} \\
\hline
Equilibrium distance
& \multicolumn{2}{c|}{$z_{\mathrm{eq}}=2.35~\text{\AA}$} \\
\hline
Electrostatic:
& \multicolumn{2}{c|}{} \\
\hline
Capacitance model
& \multicolumn{2}{c|}{Hudlet model with cantilever contribution} \\
\hline
Vacuum permittivity
& \multicolumn{2}{c|}{$\epsilon_0=8.8541878128\times10^{-12}~\mathrm{F\,m^{-1}}$} \\
\hline
Bias-modulation frequency
& \multicolumn{2}{c|}{$f_{\mathrm{mod}}=f_2-f_1=f_{2,0}-f_{1,0}=795~\mathrm{kHz}$} \\
\hline
DC bias
& \multicolumn{2}{c|}{$V_{\mathrm{DC}}=0~\mathrm{V}$} \\
\hline
CPD
& \multicolumn{2}{c|}{$V_{\mathrm{cpd}}=+100~\mathrm{mV}$} \\
\hline
Bias-modulation depth
& \multicolumn{2}{c|}{$U_{\mathrm{mod}}=100~\mathrm{mV}$} \\
\hline
Bias-modulation phase
& \multicolumn{2}{c|}{$\Phi_{\mathrm{mod}}=0$} \\
\hline

\multicolumn{3}{|l|}{\textbf{Time \& frequency sampling:}} \\
\hline
Sampling frequency
& \multicolumn{2}{c|}{$f_{\mathrm{samp}}=10~\mathrm{MHz}$} \\
\hline
Sampling period
& \multicolumn{2}{c|}{$T_{\mathrm{samp}}=1/f_{\mathrm{samp}}=100~\mathrm{ns}$} \\
\hline
Analysis-window duration
& \multicolumn{2}{c|}{$T_w=0.1~\mathrm{s}$} \\
\hline
Spectral resolution
& \multicolumn{2}{c|}{$\delta f=1/T_w=10~\mathrm{Hz}$} \\
\hline
Number of samples
& \multicolumn{2}{c|}{$N_{\mathrm{samp}}=f_{\mathrm{samp}}T_w=10^6$} \\
\hline

\multicolumn{3}{|l|}{\textbf{$z$ sampling:}} \\
\hline
Lower bound of the $z$-grid
& \multicolumn{2}{c|}{$z_{\mathrm{grid}}^{\min}=2~\text{\AA}$} \\
\hline
Upper bound of the $z$-grid
& \multicolumn{2}{c|}{$z_{\mathrm{grid}}^{\max}=20~\mathrm{nm}$} \\
\hline
$z$ sampling
& \multicolumn{2}{c|}{logarithmic grid} \\
\hline
Number of samples
& \multicolumn{2}{c|}{$N_z=9800$} \\
\hline
\end{tabular}
\end{adjustbox}
\end{table}

\begin{table}[htbp]
\centering
\caption{Time-domain reconstruction errors of the truncated Taylor-series description relative to the exact CG signal. The errors are evaluated over one first-eigenmode period, $T_1$, in the monomodal case and one super-period, $T_s$, in the bimodal case. The normalized errors are expressed relative to the peak-to-peak variation of the exact signal over the corresponding time window.}
\label{tab:time_reconstruction_errors}
\scriptsize
\renewcommand{\arraystretch}{1.2}
\begin{tabular}{|c|c|c|c|c|c|c|}
\hline
\textbf{Regime}
&
\makecell{\textbf{Taylor}\\\textbf{order}}
&
\makecell{\textbf{Time}\\\textbf{window}}
&
\makecell{\textbf{RMSE}\\$[\mathrm{nF/m}]$}
&
\makecell{\textbf{Maximum error}\\$[\mathrm{nF/m}]$}
&
\makecell{\textbf{Normalized}\\\textbf{RMSE}}
&
\makecell{\textbf{Normalized}\\\textbf{maximum error}}
\\
\hline

\multirow{2}{*}{Monomodal}
&
$3$
&
$T_1$
&
$1.00\times10^{-2}$
&
$2.83\times10^{-2}$
&
$6.80\%$
&
$19.2\%$
\\
&
$30$
&
$T_1$
&
$1.99\times10^{-8}$
&
$8.74\times10^{-8}$
&
$1.30\times10^{-5}\%$
&
$5.90\times10^{-5}\%$
\\
\hline

\multirow{2}{*}{Bimodal}
&
$3$
&
$T_s$
&
$1.16\times10^{-2}$
&
$4.87\times10^{-2}$
&
$6.28\%$
&
$26.4\%$
\\
&
$30$
&
$T_s$
&
$1.88\times10^{-7}$
&
$\approx 2.0\times10^{-6}$
&
$1.02\times10^{-4}\%$
&
$9.75\times10^{-4}\%$
\\
\hline

\end{tabular}
\end{table}

\clearpage

\end{document}